\documentclass[twocolumn,superscriptaddress]{revtex4}
\usepackage{latexsym, amsmath, amssymb,amsthm,amsopn,amsfonts, amsbsy, graphicx, epstopdf, multirow}
\usepackage{breqn}
\usepackage{color}
\usepackage{ulem}

\def\botimes{\boldsymbol{\otimes}}
\def\b2otimes{\boldsymbol{\bar\otimes}}

\newcommand{\ukarc}{\ensuremath{\mu}K-arcmin}

\def\simlt{\lower.5ex\hbox{$\; \buildrel < \over \sim \;$}}
\def\simgt{\lower.5ex\hbox{$\; \buildrel > \over \sim \;$}}
\def\simgtalt{\lower.5ex\hbox{$\buildrel > \over \sim \;$}}

\def\b#1{{\bar{#1}}}

\def\bd#1{{\bf #1}}
\def\l#1{\left #1}
\def\r#1{\right #1}

\def\eref#1{(\ref{#1})}

\makeatletter
\let\cat@comma@active\@empty
\makeatother

\begin{document}
\title{Forecasting performance of CMB experiments in the presence of complex foreground contaminations.}

\author{Radek Stompor\footnote{E-mail: radek@apc.univ-paris-diderot.fr}}
\affiliation{AstroParticule et Cosmologie, Univ Paris Diderot, CNRS/IN2P3,CEA/Irfu, Obs de Paris, Sorbonne Paris Cit\'e, France}
\author{Josquin Errard\footnote{E-mail: josquin.errard@lpnhe.in2p3.fr}}
\affiliation{Sorbonne Universit\'es Institut Lagrange de Paris (ILP), 98 bis Boulevard Arago 75014 Paris, France }
\affiliation{LPNHE, CNRS-IN2P3 and Universit\'es Paris 6 \& 7, 4 place Jussieu F-75252 Paris, Cedex 05, France}
\author{Davide Poletti\footnote{E-mail: Davide.Poletti@apc.univ-paris7.fr}}
\affiliation{AstroParticule et Cosmologie, Univ Paris Diderot, CNRS/IN2P3,CEA/Irfu, Obs de Paris, Sorbonne Paris Cit\'e, France}

\date\today

\begin{abstract}
We present a new, semi-analytic framework for estimating the level of residuals present in CMB maps derived from multi-frequency
Cosmic Microwave Background (CMB) data and forecasting their impact on cosmological parameters. 
The data are assumed to
contain non-negligible signals of astrophysical and/or Galactic origin, which we clean
using parametric component separation technique. We account for discrepancies between
the foreground model assumed during the separation procedure and the true one, allowing for differences in scaling
laws and/or their spatial variations. 
Our estimates and their uncertainties include both systematic and statistical effects and are averaged over the instrumental noise and CMB signal realizations.
The framework can be further extended to  account self-consistently for existing uncertainties in the foreground
models. 
We  demonstrate and validate the framework on simple study cases which aim at estimating the tensor-to-scalar ratio, $r$. 
The proposed approach is computationally efficient permitting an investigation of hundreds of set-ups and foreground models on a single CPU.
\end{abstract}

\maketitle

\section{Introduction.}
Forecasting performance of current and future CMB experiments is a necessary step in conception, design and optimization of their
hardware as well as operations.
Ideally, a forecasting procedure should be both reliable and efficient permitting  scrutiny of broad swaths of parameter space in order to quickly zoom on
a limited subset of the most promising configurations. This subset  should be small enough to facilitate their further, more detailed investigation, 
typically employing  numerical simulations, which while permitting  a higher level of realism and detail are significantly more time and resource consuming.

Reliable forecasting for high precision CMB experiments is difficult due to the presence of the non-CMB signals, which unavoidably contribute
to the measurements registered by the CMB instruments.
Indeed, the multi-frequency observations from the Planck and WMAP satellites indicate that foreground emissions originating from our Galaxy or extra-galactic sources
represent a major contaminant, e.g.,~\cite{Gold2011, PlanckXX2015,Krach2016}, which current and future CMB polarization experiments will have to deal explicitly with. Methods employed for this purpose will thus have to ensure precision matching sensitivity
envisaged for these forthcoming efforts and set by very ambitious science goals, which the CMB community world-wide is preparing to address.
These goals include a detection and a characterization of the B-mode signal over a broad range of angular scales with a special emphasis on its large angular scale part, which is thought to be generated by primordial gravity waves
present in the early Universe. The key parameter in this latter case is the so-called tensor-to-scalar ratio, $r$, and for concreteness in the following we
will couch our presentation as targeting constraints on this parameter. The approach we introduce is however fully general and generalization to 
other parameters is  straightforward. 

The standard CMB forecasting tools are  ill-adapted to tackle cases with non-negligible foreground contributions. Their impact 
is therefore often either modelled or assessed by some simplified means either in respect to estimating the residuals or their
impact on the detection of $r$, e.g.,~\cite{Snowmass2015a, Verde2006,  Amblard2007, dunkley2009, bonaldi2011, creminelli2015, Krach2016, Kogut2016}. Alternately, the issue is investigated with help of numerical simulations, which are typically computationally
heavy and thus only  allow for a limited number of studied cases~\cite{Katayama2011,Remazeilles2016, Alonso2016}.

Against this background, Errard et al (2011)~\cite{Errard2011} has proposed a semi-analytic framework, which 
attempts to propagate strictly statistical uncertainties incurred as a result of a component separation procedure
to the final estimate  of $r$. Their component separation of choice is a maximum likelihood parametric component separation approach~\cite{brandt1994, 2006ApJ...641..665E, Stompor2009}, which
assumes a parametrization of the frequency scalings for  each considered sky component.
Though self-consistent this approach is only capable of dealing with the statistical uncertainties and therefore its conclusions are limited
in their validity and the results should be interpreted with caution. Specifically, this approach requires that the parametrization
assumed for the frequency scaling of the sky components is sufficiently flexible and general that the actual frequency scaling laws of the true sky signals is included as its special case.
Nonetheless, the framework has proven to be helpful in enabling studies of numerous experimental set-ups in a uniform fashion~\cite{Errard2012, Errard2015}, providing
useful insights and intuitions and informing multiple instrument designs.

In this work we develop a framework capable of accounting for differences between these two sky signal models. As in~\cite{Errard2011}, we assume that the components are separated with help of the parametric component separation technique and we estimate
by semi-analytic means both the bias and statistical uncertainty, which are both present whenever the two sky models do not match.
This can be either due to differences in the frequency scaling laws for some of the components or their spatial variability. The framework 
also permits incorporating the uncertainty related to our ignorance of the foreground signals and/or shortcomings of our models.

The bias and statistical uncertainties are then propagated to the second step of the procedure, where their impact on $r$ is calculated. The new approach is equivalent to that of Errard et al (2011)~\cite{Errard2011}, if the sky model and the true sky are consistent, and in this sense it extends and completes this earlier work.

We present the formalism in Sect.~\ref{sect:Framework} and demonstrate and validate it in Sect.~\ref{sect:applic}. We leave a thorough 
investigation of different experimental set-ups and foreground models for  future work. For convenience, we define symbols most commonly used in this paper in Table I.

\begin{table*}[htb!]
\caption{Notations}
\centering
\label{table:notations}
\resizebox{15cm}{!}{
\hspace{-2cm}
\begin{tabular}{c|c|c|c|c|c|c|c|c|c|c|c|c|c} 
\hline
\hline
symbols  & $\mathbf{d}$ &  $\mathbf{\hat{s}}$ & $\bd{n}$, $\bd{N}$ & $\bd{\hat d}$ & $\mathbf{\hat A}$ & $p$ & $k$ &$\beta$ & $\mathbf{A}$ & $\mathbf{s}$ & $\mathbf{\bar m}_p$ & $\mathbf{r}_p$, $\mathbf{r}^{\rm cmb} (\beta)$, ${C}^{\rm res}_{\ell}$& $\mathbf{\hat f}_p$, $\mathbf{F}_{p k}$, ${\cal F}_\ell^{\rm fore} $\\
\hline
\multirow{3}{*}{definition} & set of observed &  true sky  & noise, & true sky & true mixing & sky& frequency & spectral & model &  model sky  & noiseless  & noiseless & foregrounds  \\
 & multi-frequency &noiseless & covariance & component & matrix &pixel &channel & parameters & mixing &component & estimates of  & residuals & signal \\
 & maps & signal & & amplitudes &  & & & &matrix & amplitudes &the components &  &  \\
\hline
\hline
\end{tabular}}
\end{table*}

\section{Framework}

\label{sect:Framework}

\subsection{Data model.}

The outline of our approach is as follows. Our input data are assumed to consist of a set of multi-frequency maps. These are collated together in a single data vector, $\mathbf{d}$, 
and are assumed to be a linear combination of sky component amplitudes in corresponding sky pixels. Collecting these together in a single sky component vector, $\mathbf{\hat{s}}$, we can therefore write,
\begin{eqnarray}
\bd{d} = \bd{\hat d} + \bd{n} \equiv \bd{\hat{A}}\,\bd{\hat{s}} \, + \, \bd{n},
\end{eqnarray}
where $\bd{\hat d}$ denotes true sky (noiseless) signal, matrix $\mathbf{\hat A}$ stands for the true mixing matrix of our data, and $\mathbf{n}$ -- noise. 
We therefore have for a specific pixel, $p$,
\begin{eqnarray}
\bd{d}_p & = & \bd{\hat d}_p + \bd{n}_p \equiv \bd{\hat{A}}_p\,\bd{\hat{s}}_p \, + \, \bd{n}_p,
\label{eqn:trueSkyModel}
\end{eqnarray}
where $\bd{d}_p$ stands for a vector of sky signal amplitudes measured at all observed frequencies in pixel $p$. Similarly,
for a channel, $k$,
\begin{eqnarray}
\bd{d}^{\l(k\r)} & = & \bd{\hat d}^{\l(k\r)} + \bd{n}^{\l(k\r)} \equiv \bd{\hat{A}}^{\l(k\r)}\,\bd{\hat{s}}^{\l(k\r)} \, + \, \bd{n}^{\l(k\r)},
\end{eqnarray}
where $\bd{d}^{\l(k\r)}$ is a single frequency map of the observed sky in a frequency band defined by $k$.
 
We assume hereafter that the actual mixing matrix, $\mathbf{\hat A}$, is not available to us and instead we have to rely on some model of it
to represent the available data. We denote this assumed mixing matrix as $\mathbf{A}$ and the corresponding component vector as $\mathbf{s}$. 
Our assumed data model therefore states that,
\begin{eqnarray}
\bd{d}_p & = & \bd{A}_p(\beta)\,\bd{s}_p \, + \, \bd{n}_p,
\label{eqn:dataModel}
\end{eqnarray}
where $\beta$ denotes parameters used to parametrize the mixing matrix in order to reduce the number of unknowns.

Hereafter, we allow for a pixel-dependence of the model mixing matrix, $\mathbf{A}$. This could be either due to allowing for different values of the same physical parameters in different subsets of pixels 
or due to adopting different physical laws and parametrizations in different pixels. In any case, hearafter, a parameter, which is allowed to have a  different value in two different pixels is treated as two different parameters rather than as a single parameter, which is pixel-dependent. This perspective will be helpful in the following.  

Note that, whenever sky-variability of the scaling parameters is considered later on, we will assign different parameters to subsets of all pixels and make
an implicit assumption hereafter that these are composed of a rather large number of pixels, covering well-behaved, singly-connected, compact 
sky patches. Though, the formalism is applicable more generally, its implementation and the interpretation of its results are both aided by this assumption.

We emphasize that in the presented formalism no assumption is made about the true sky signals and mixing matrices. Though, these are obviously
needed for any specific application of the formalism in order to define the statistical properties of the data and play a crucial role in determining the resulting forecasts.
\subsection{Data likelihood.}

Following our data model we can write the standard likelihood function for the data given our model, Eq.~\eref{eqn:dataModel}. This reads up to an irrelevant constant as,
\begin{eqnarray}
{\cal S}_{map} = \sum_p\,(\bd{d}_p-\bd{A}_p(\beta)\,\bd{s}_p)^t \,\bd{N}_p^{-1} (\bd{d}_p-\bd{A}_p(\beta)\,\bd{s}_p),
\label{eqn:likeMapFull}
\end{eqnarray}
where here and in the following we define ${\cal S}$ as equal to $-2 \ln {\cal L}$ up to some constant. We note that in the
above expression the noise covariance is explicitly assumed to be uncorrelated between pixels. This is clearly not always the case and may need to be taken into account (see, e.g., a relevant discussion in~\cite{Alonso2016}). Mathematically, the formalism presented hereafter
can be easily generalized to permit correlations between pixels, however, in actual implementation treating full pixel-pixel covariance matrices becomes
quickly prohibitively expensive. 
As our goal here is to provide a quick, performance forecasting tool, we will thus neglect the potential presence of 
such correlations.

If the multifrequency maps do not conform with our assumed model, i.e., $\mathbf{A} \ne \mathbf{\hat A}$,
this likelihood is obviously incorrect as
there is no value of $\beta$ for which $\bd{d}_p-\bd{A}_p(\beta)\bd{s}_p$ could be merely a noise. However, this is the likelihood we would have adopted for component
separation in the absence of any other information about the true sky.
If  discrepancies between the assumed and true sky signal models are present, we expect that relying on this likelihood will lead in general to both systematic and statistical uncertainties in the derived results and in particular
to systematic and statistical foreground residuals in the separated CMB map. In the approach proposed here we aim at estimating both these residuals and evaluating their impact on a value of the tensor-to-scalar ratio parameter, $r$,
derived from the separated CMB map. 

The proposed procedure involves two main steps: component separation and parameter estimation, which we describe in  detail in the following.

\subsection{Parametric component separation.}

\label{ssect:compSep}

\subsubsection{Spectral parameters}

Following~\cite{Stompor2009, Errard2011} we invoke the spectral likelihood, which we will use to determine the spectral parameters of the scaling relations,
\begin{eqnarray}
{\cal S}_{spec}\hskip -2pt=\hskip -2pt -\hskip -2pt\sum_{p}\l(\bd{A}_p\bd{N}_p^{-1}\bd{d}_p\r)^t\l( \bd{A}_p^t\bd{N}_p^{-1}\bd{A}_p\r)^{-1}\bd{A}_p\bd{N}_p^{-1}\bd{d}_p.
\label{eqn:profileLikeDef}
\end{eqnarray} 
This is a profile likelihood obtained by maximizing the map likelihood in Eq.~\eref{eqn:likeMapFull} with respect to the sky signal, $\mathbf{s}$. As such
it peaks at exactly the same values as the full likelihood. ${\cal S}_{spec}$ can be maximized case-by-case for any given data, what indeed is implicitly
or explicitly done in the parametric component separation codes, e.g.~\cite{2006ApJ...641..665E}, or, instead, first averaged over the statistical ensemble of plausible input data
and then maximized over the spectral parameters to yield both their average estimate and error on them. In the simplest case the statistical ensemble 
can merely include realizations of the noise. However, more generally, instead of a single true model of the foreground signals we may prefer to
consider a family of models defined by their stochastic properties. This can be either due to our imperfect understanding of the 
foreground physics or due to the actual complexity of the foreground, which may be easier to sum up by statistical means~\cite{poh2016, caldwell2016}.
 The formalism presented
here lends itself straightforwardly to this kind of extensions. Nonetheless, we leave their exploitation to future work and in this paper we focus on
the spectral likelihood averaged over a statistical ensemble of the noise realizations, which is then given by, 
\begin{eqnarray}
\langle {\cal S}_{spec} \rangle =  -{\rm tr}\, \sum_p \bigg\{(\bd{N}_p^{-1} - \bd{P}_p) \Big( \bd{\hat d}_p \bd{\hat d}_p^t+ \bd{N}_p\Big)\bigg\}.
\label{eqn:likeAvFinal}
\end{eqnarray}
 Here, the dependence on the spectral parameters is confined to the projection operator, $\mathbf{P}_p$, 
 \begin{eqnarray}
\bd{P}_p \equiv \bd{N}_p^{-1} - \bd{N}_p^{-1}\bd{A}_p\l(\bd{A}_p^t\bd{N}_p^{-1}\bd{A}_p\r)^{-1} \bd{A}_p^t \bd{N}_p^{-1}.
\label{eqn:projDef}
 \end{eqnarray}
This likelihood can be maximized very efficiently numerically, given that in most applications the number of unknown spectral parameters is rather limited and capitalizing on 
the analytical derivatives of the likelihood as derived in Appendix~\ref{app:mapLike}. In the approach proposed here, these maximum likelihood values define the average values of the spectral 
parameters as could be derived from the actual data while the curvature of the likelihood computed at its peak quantifies the uncertainty expected due to the instrumental noise. Denoting the latter as $\boldsymbol{\Sigma}$, we  have,
\begin{eqnarray}
\left({\boldsymbol \Sigma}^{-1}\right)_{\beta\beta'} \equiv \left\langle \frac{\partial^2{\cal S}}{\partial \beta \partial \beta'}\right\rangle.
\label{eqn:specParError}
\end{eqnarray}
We note that the proposed procedure can be always applied, whether the assumed and true sky models match or not.
When the true mixing matrix, $\mathbf{\hat A}$, agrees with the assumed one, $\mathbf{A}(\beta)$, for some values of the parameters, $\beta$, 
then the estimated values agree with these and the estimator is unbiased. In this case, only statistical residual, related to the statistical scatter of the determined values of $\beta$ due to the instrumental noise, is present in the cleaned CMB map, which, if properly accounted for, will merely increase the statistical uncertainty of subsequently estimated cosmological parameters, without biasing their values~\cite{Errard2011}.

When the assumed and true mixing matrices do not
perfectly coincide for any values of the parameters for some or all pixels, be it due the inconsistency of the scaling laws or their spatial variability or both, there is a systematic residual unavoidably present in the estimated CMB map, which in turn may lead to biases in the estimated values of the cosmological parameters. 
We note that in such cases there are in general no 'true' values of the spectral parameters and their estimated, 'effective' values depend not only on the assumed and true mixing matrices but also instrument characteristics such as observational frequency bands, etc.

 The statistical error matrix, $\boldsymbol{\Sigma}$, can be computed semi-analytically using Eq.~\eref{eqn:profileLikeSecondDer}, or numerically. Indeed, many maximization routines construct a numerical approximation
to the curvature matrix as part of the procedure. We have found that in our test cases both approaches produced results in very good agreement.
We note that in the absence of calibration errors and assuming that the spectral dependence of the CMB component is known completely, neither the estimated spectral parameter values nor their uncertainties depend on the CMB signal present in the data. This observation is analogous to the one pointed out in~\cite{Errard2011} and is elaborated on in Appendix~\ref{app:mapLike}.

\subsubsection{Residuals}

\label{s2sect:compSepRes}

Given $\mathbf{A}_p$ computed for some value of $\beta$, we can express noiseless estimates of the components as~\cite{Stivoli2010},
\begin{eqnarray}
\mathbf{\bar m}_p = (\mathbf{A}_p^t\mathbf{N}_p^{-1}\mathbf{A}_p)^{-1}\mathbf{A}_p^t\mathbf{N}_p^{-1}\mathbf{\hat d}_p \equiv \mathbf{W}_p(\beta)\mathbf{\hat d}_p.
\label{eqn:compSepEq}
\end{eqnarray}
The noiseless residuals in the estimates then read as,
\begin{eqnarray}
\mathbf{r}_p = \mathbf{\bar m}_p - \mathbf{\hat s}_p = \mathbf{W}_p(\beta)\mathbf{\hat d}_p - \mathbf{\hat s}_p.
\label{eqn:resPixDefGen}
\end{eqnarray}
Hereafter, we assume that the CMB corresponds to the $0$th element of any multicomponent vector and split the multi-component vectors and the mixing matrices into  CMB and foreground parts as follows,
\begin{eqnarray}
\mathbf{\hat A} \hskip -1pt \equiv  \hskip -1pt [\mathbf{\hat A}^{\rm cmb}, \mathbf{\hat A}^{\rm fore}], \ \mathbf{A}  \hskip -1pt \equiv  \hskip -1pt [\mathbf{ A}^{\rm cmb}, \mathbf{A}^{\rm fore}],
\ \mathbf{\hat s}_p  \hskip -1pt \equiv  \hskip -1pt
\left[ 
\begin{array}{c}
\mathbf{\hat s}^{\rm cmb}\\
\mathbf{\hat s}^{\rm fore}
\end{array}\right]\hskip -4pt . \ \ 
\end{eqnarray}
We can now represent $\mathbf{\hat d}_p$ as,
\begin{eqnarray}
\mathbf{\hat d}_p = \mathbf{\hat A}^{\rm cmb}_p \mathbf{\hat s}_p^{\rm cmb} + \mathbf{\hat A}^{\rm fore}_p\mathbf{\hat s}_p^{\rm fore} \equiv 
\mathbf{\hat A}^{\rm cmb}_p \mathbf{\hat s}_p^{\rm cmb} + \mathbf{\hat f}_p, 
\label{eqn:trueDataSplit}
\end{eqnarray}
where $\mathbf{\hat f}$ denotes a true noise-free contribution of the foregrounds to all single-frequency maps.
As we assume throughout that the CMB scaling is the same in the model and true sky, we have,
\begin{eqnarray}
\left[\mathbf{W}_p\,\mathbf{\hat A}^{\rm cmb}\right]_{00} = 1.
\label{eqn:waEqn}
\end{eqnarray}
This emphasizes the fact that all of the CMB signal will remain in the estimated CMB  component, which
however will be contaminated by contributions from the other, non-CMB signals.

Indeed, on rewriting Eq.~\eref{eqn:resPixDefGen} and specializing it for the CMB component residual only we have,
\begin{eqnarray}
\mathbf{r}_p^{\rm cmb} = \sum_k \mathbf{W}^{0 k}_p(\beta)\mathbf{\hat f}_{p}^{\l(k\r)} \equiv \sum_k \mathbf{W}^{0 k}_p(\beta)\mathbf{F}_{p k} 
\label{eqn:resCMBpix}
\end{eqnarray}
which, as expected, is explicitly free of the CMB signal. Here, we have introduced a foreground matrix, $\mathbf{F}$, $k$th column of which defines the
total foreground contribution to the $k$th frequency channel.

We can now perform
a Taylor expansion of the residuals with respect to the scaling parameters but around their estimated, maximum likelihood values, $\bar \beta$, obtaining,
\begin{eqnarray}
\begin{array}{l l l}
\mathbf{r}^{\rm cmb}_{p}(\beta)  & \simeq & {\displaystyle \sum_k \mathbf{W}^{0 k}_p(\bar \beta)\mathbf{F}_{pk}
+ \sum_{k, \beta} \,\delta \beta \left.\frac{\partial \mathbf{W}^{0k}_p}{\partial \beta}\right|_{\bar \beta}\mathbf{F}_{pk}}\\
&+& {\displaystyle \sum_{k, \beta,\beta'} \,\delta \beta \delta \beta' \left.\frac{\partial^2 \mathbf{W}^{0k}_p}{\partial \beta\partial \beta'}\right|_{\bar \beta}\mathbf{F}_{pk}},
\end{array}
\label{eqn:resExp}
\end{eqnarray}
where we need to go up to the second order to have a consistent, up to the second order, approximation of the data covariance matrix, $\mathbf{E}$, in Eq.~\eqref{eqn:EmatDef}.
On introducing pixel-domain objects: a vector, $\mathbf{y},$ two dimensional, $\mathbf{Y}^{\l(1\r)}$, and three dimensional, $\mathbf{Y}^{\l(2\r)}$, arrays, defined as,
\begin{eqnarray}
\mathbf{y}_p &\equiv& {\displaystyle  \sum_k \mathbf{W}^{0 k}_p(\bar \beta)\,\mathbf{F}_{pk},}\nonumber\\
\mathbf{Y}^{\l(1\r)}_{p \beta} & \equiv &  {\displaystyle \sum_k \left.\frac{\partial \mathbf{W}^{0k}_p}{\partial \beta}\right|_{\bar \beta}\mathbf{F}_{pk},}
\label{eqn:yVectsDef}\\
\mathbf{Y}^{\l(2\r)}_{p \beta\beta'} & \equiv &  {\displaystyle \sum_k \left.\frac{\partial^2 \mathbf{W}^{0k}_p}{\partial \beta\partial \beta'}\right|_{\bar \beta}\mathbf{F}_{pk},}\nonumber
\end{eqnarray}
we can rewrite this last expression as,
\begin{eqnarray}
\mathbf{r}^{\rm cmb} (\beta) \equiv  \mathbf{y} + \sum_\beta \delta\beta\, \mathbf{Y}^{\l(1\r)}_{\beta} + \sum_{\beta, \beta'} \delta\beta \delta\beta' \, \mathbf{Y}^{\l(2\r)}_{\beta \beta'},
\label{eqn:resTaylor2}
\end{eqnarray}
where for shortness we use $\mathbf{Y}^{\l(1\r)}_{\beta}$ and $\mathbf{Y}^{\l(2\r)}_{\beta \beta'}$ to denote pixel-domain vectors given by the elements of the arrays, $\mathbf{Y}^{\l(1\r)}$ 
and $\mathbf{Y}^{\l(2\r)}$ for which the spectral parameter indices -- $\beta$ and $\beta, \beta'$ -- are fixed.

We point out that $\delta \beta$ is  explicitly pixel-independent. This is so thanks to the way we define the total spectral parameter set as discussed following Eq.~\eref{eqn:dataModel}, where every parameter appears as
many times in the parameter set as many independent values it is allowed to take.
Clearly, not all the parameters defined in this way will in general be relevant for all pixels. This is encoded in the multi-dimensional arrays, $\mathbf{Y}^{\l(1\r)}$ and $\mathbf{Y}^{\l(2\r)}$, which will have all entries corresponding
to such pixels set to zero.

Eq.~\eqref{eqn:resTaylor2} is a generalization of Eq. (10) of Errard et al. (2011) and Eq.~(24) of Stivoli et al. (2010)~\cite{Stivoli2010}.
The generalization concerns two aspects:
\begin{enumerate}
\item first,  it includes the bias in the estimated component maps due to the residual foregrounds, 
which does not disappear when averaged over the statistical ensemble of noise and foreground realizations. This is given by $\mathbf{y}$. 
\item second, this equation has been derived in a way, which did not invoke any assumptions about the
pixel-dependence of the true sky mixing matrix, $\mathbf{\hat A}_p$, which can be therefore arbitrary.
\end{enumerate}

Though this expression is derived in the pixel domain we can rewrite it in the harmonic domain owing to the fact that $\delta \beta$ are pixel independent,
\begin{eqnarray}
\mathbf{\tilde r}^{\rm cmb} (\beta) \equiv  \mathbf{\tilde y} + \sum_\beta \delta\beta\, \mathbf{\tilde Y}^{\l(1\r)}_{\beta} + \sum_{\beta, \beta'} \delta\beta \delta\beta' \, \mathbf{\tilde Y}^{\l(2\r)}_{\beta \beta'},
\label{eqn:resTaylorHar}
\end{eqnarray}
where we use a tilde to denote vectors of harmonic multipoles.
 For definiteness they are arranged in such a way that the multipoles with the same $\ell$ are ordered consecutively
with $m$ increasing from $-\ell$ to $\ell$ and are followed by the modes with $\ell' = \ell+1$. So the relations between the multipole numbers, $(\ell, m)$, 
of a multipole and its position, $j$, in the harmonic vector are as follows,
\begin{eqnarray}
j & = & \ell^2+\ell+m, \nonumber \\
 \ell & =  & {\rm round}[(-1 + \sqrt{ 1 + 4 j })/2],\label{eqn:j2lm}\\
 m & = & j - \ell \l(\ell+2\r), \nonumber
\end{eqnarray}
where $j$ goes from $0$ to $(\ell_{max}+1)^2-1$ and the function ${round}$ rounds a real number to the {\it closest} integer. 

We note that going from the pixel domain vectors, $\mathbf{y}, \mathbf{Y}^{\l(1\r)}_{\beta}, \mathbf{Y}^{\l(2\r)}_{\beta\beta'}$, to their harmonic domain
counterparts,  $\mathbf{\tilde y}, \mathbf{\tilde Y}^{\l(1\r)}_{\beta}, \mathbf{\tilde Y}^{\l(2\r)}_{\beta\beta'}$, is only straightforward, if the pixel-domain vectors corresponds
to the full-sky maps. This obviously is rarely the case in practice, as even for the full sky observations, the
parts most affected by the foregrounds, i.e., Galactic plane and point sources, need to be typically masked out and are not used for the component
separation.
 In the spirit of the Fisher approaches we however ignore this difficulty hereafter, assuming that this is
possible and merely comes at the cost of the increased statistical uncertainty due to a fewer number of available modes and some rough
cut-off scale at low-$\ell$. In practice, as mentioned earlier, see the discussion after Eq.~(\ref{eqn:dataModel}), this implies that the adopted mixing matrix 
instead of being permitted to change freely from pixel-to-pixel is taken to be the same for sufficiently large and regular sky patches.

For future convenience, we can write Eq.~\eref{eqn:resTaylorHar} for each harmonic mode, $j\;(=\ell^2+\ell+m$), Eqs.~(\ref{eqn:j2lm}), as,
\begin{eqnarray}
\mathbf{\tilde r}_j^{\rm cmb} (\beta) \equiv  \mathbf{\tilde y}_j +  \mathbf{\tilde Y}^{\left(1\right)}_{j} {\boldsymbol \delta}
+ {\boldsymbol \delta}^t \mathbf{\tilde Y}^{\left(2\right)}_j {\boldsymbol \delta},
\label{eqn:resTaylorHarFinal}
\end{eqnarray}
where $\mathbf{\tilde Y}^{\left(1\right)}_{j}$ and $\mathbf{\tilde Y}^{\left(2\right)}_j$ stand for a vector and a matrix respectively
made of elements of $\mathbf{\tilde Y}^{\left(1\right)}_{j\beta}$ and $\mathbf{\tilde Y}^{\left(2\right)}_{j\beta\beta'}$ for the given $j$
 and ${\boldsymbol \delta}$ is a vector of uncertainties on spectral parameters, $\beta$, around the estimated values of the parameters. 
 We therefore have
 \begin{eqnarray}
 \langle {\boldsymbol \delta} {\boldsymbol \delta}^t \rangle = { \boldsymbol \Sigma} \ \ \ \hbox{\rm and} \ \ \ \langle\boldsymbol{\delta}\rangle = 0.
 \end{eqnarray}
Using these two last equations we can now rewrite an expression for the typical level of the residuals in the power
spectrum domain. Indeed, we have, (see Appendix~\ref{app:psRes} for details),
\begin{eqnarray}
{C}^{\rm res}_{\ell} 
 & \simeq & \botimes_\ell(\mathbf{\tilde y},  \mathbf{\tilde y}) + \botimes_\ell( \mathbf{\tilde y}, \mathbf{\tilde z}) + \botimes_\ell(\mathbf{\tilde z}, \mathbf{\tilde y}) 
       \label{eqn:psResFinal}
 \ \ \ \ \ \ \ \ \\
      & &\ \ \ \ \ \ \ \ \ \ \ +  {\rm tr} \Big[ \boldsymbol{\Sigma} \,  \botimes_\ell\hskip -2pt(\mathbf{\tilde Y}^{\l(1\r)},  \mathbf{\tilde Y}^{\l(1\r)})\Big],\nonumber
\end{eqnarray}
where $\mathbf{\tilde z}$ is defined as,
\begin{eqnarray}
\mathbf{\tilde z}_j \equiv {\rm tr}\, \Big[ \mathbf{\tilde Y}_j^{\l(2\r)}\boldsymbol{\Sigma}\Big],
\end{eqnarray}
and can be computed as the harmonic representation of the pixel domain object, $\mathbf{z}$, defined as, 
\begin{eqnarray}
\mathbf{z}_p & = & {\rm tr} \l[ \mathbf{Y}^{\left(2\right)}_{p} {\boldsymbol \Sigma}\r]
= \sum_{\beta,\beta'} \sum_k \left.\frac{\partial^2 \mathbf{W}^{0k}_p}{\partial \beta\partial \beta'}\right|_{\bar \beta}\mathbf{F}_{p k}\, {\boldsymbol \Sigma}_{\beta' \beta}\nonumber \\
& = & \sum_k  \mathbf{F}_{p k} \l[\sum_{\beta,\beta'} \l( \left.\frac{\partial^2 \mathbf{W}^{0k}_p}{\partial \beta\partial \beta'}\right|_{\bar \beta} {\boldsymbol \Sigma}_{\beta' \beta}\r)\r].
\label{eqn:zFullPix}
\end{eqnarray}
In addition, we have also introduced symbol $\botimes$ to denote a power spectrum of two sets of harmonic coefficients provided as input parameters, i.e.,
\begin{eqnarray}
\botimes_\ell(\mathbf{\tilde X},  \mathbf{\tilde Z}) \equiv \frac{1}{2\ell+1}\,\sum_{m = -\ell}^{\ell} \mathbf{\tilde X}_j^\dagger \mathbf{\tilde Z}_j, \ \ j = \ell^2+\ell+m.\ \ 
\label{eqn:xspecDef}
\end{eqnarray}
We note that whenever, $\mathbf{\tilde X}$ and  $\mathbf{\tilde Z}$ are multi-dimensional arrays of spectral coefficients indexed by $j$ and the outcome of the operations is a 
matrix containing $\ell$th multipole of all the (cross)spectra of all these coefficients.

Eq.~\eref{eqn:psResFinal} permits computation of the
power spectra of the typical residual in the presence of: (1) spatial variability of the spectral indices, in both the model and the actual sky, (2) discrepancies
in the spatial and frequency behavior between the two, and (3) inhomogeneity of the measurement noise, which is however assumed to be uncorrelated between pixels.
The inputs required for this are the relevant auto- and cross- spectra of $\mathbf{\tilde y}, \mathbf{\tilde Y}^{\l(1\r)}$, and $\mathbf{\tilde z}$.
These spectra, in general, do not have any simple physical interpretation as they conflate
all the different pixel-dependent effects together. However, if the assumed sky model is pixel-independent and the noise is white, all the spectra can be directly related to the
spectra of the combined foreground signals as measured in different frequency bands, Appendix~\ref{app:simpCase}, i.e.,
\begin{eqnarray}
{\cal F}_\ell^{\rm fore} \equiv \botimes_\ell( \mathbf{\tilde F}, \mathbf{\tilde F}),
\label{eqn:bandXspecDef}
\end{eqnarray}
where $\mathbf{\tilde F}$ is a harmonic representation of $\mathbf{F}$ defined in Eq.~(\ref{eqn:trueDataSplit}) and therefore it is a matrix of as many
columns as the assumed  frequency bands with each column representing a combined foreground signal in each frequency band. Consequently, ${\cal F}^{\rm fore}_\ell$ is then
 a matrix containing multipole $\ell$ of all auto- and cross-spectra between foreground signals in all the frequency bands.
Furthermore, if the true sky frequency scaling is also pixel independent, then the latter spectra can be related straightforwardly to those of the sky components 
at some fiducial frequency, as it was the case in~\cite{Errard2012}.

We can also compute a dispersion around the average spectrum of the residual, Appendix~\ref{app:psRes}, which can be approximated as follows,
\begin{eqnarray}
{\rm Var} (C_\ell^{\rm res}) & \simeq &
2\,\l[{\rm tr} \l(\boldsymbol{\Sigma}\,\botimes_\ell\hskip -2pt(\mathbf{\tilde Y}^{\l(1\r)},\,\mathbf{\tilde Y}^{\l(1\r)})\r)\r]^2
\label{eqn:varResFinal}
\\ & + &
2\,\botimes_\ell\hskip -2pt(\mathbf{\tilde y},  \mathbf{\tilde Y}^{\l(1\r)})\, \boldsymbol{\Sigma}\,\botimes_\ell\hskip -2pt(\mathbf{\tilde Y}^{\l(1\r)}, \mathbf{\tilde y}). \ \ \ 
\nonumber
\end{eqnarray}

\subsubsection{Noise}

\label{s2sect:compSepResNoise}

The instrumental noise present in the cleaned CMB map, $\mathbf{n}^{\rm cmb}$, is given by,
\begin{eqnarray}
\mathbf{n}^{\rm cmb}_{p} & = & {\displaystyle \sum_k \mathbf{W}^{0 k}_p(\bar \beta)\mathbf{n}_{pk}},
\end{eqnarray}
where $\mathbf{n}_{pk}$ denotes the noise of the $k$th frequency map. Given that we assume that the noise of each single frequency map is Gaussian, uncorrelated between pixels and  its pixel covariance is given by $\mathbf{N}_p$, the variance of the CMB map noise can be expressed as,
\begin{eqnarray}
\sigma^2_{\rm cmb, \, p}  & \equiv & \l[\l(\mathbf{A}^t \mathbf{N}^{-1}\mathbf{A}\r)^{-1}\r]_{\hbox{\scriptsize{\sc cmb} x {\sc cmb}}} \, \nonumber\\
& = & \, {\displaystyle \sum_{k,k'} \mathbf{W}^{0 k}_p(\bar \beta) \mathbf{W}^{0 k'}_p(\bar \beta) \mathbf{N}_{p, \, kk'}}.
\label{eqn:cmbVar}
\end{eqnarray}
The CMB map noise will be inhomogeneous and its variance -- pixel-dependent, whenever the noise of the single-frequency maps is inhomogeneous and/or the assumed scaling laws are pixel-dependent. In this latter case the coefficient of the matrix $\mathbf{W}$ will depend on the pixel, giving rise to the CMB map noise inhomogeneity even if the single-frequency maps noise is homogeneous. As mentioned earlier the formalism as developed until now is capable of handling the cases of both these kinds. Nevertheless, the noise inhomogeneity potentially leads to two problems. On the technical level, the noise which is inhomogeneous in the pixel domain results in correlations of the noise in the harmonic domain, which therefore usually can not be described in a compact, computationally-manageable manner. In particular, the noise power spectrum does not anymore provide a sufficient description of the noise properties in the harmonic domain. This is however in practice a necessary assumption for the forecasting approach as presented here. We discuss this issue in the next Section.

There is also another, more fundamental problem related to the noise inhomogeneities, which is pertinent to the robustness of the performance forecasts in the presence of the foregrounds and is not specific to this particular approach but is applicable to any forecasts based on the pixel-domain parametric approach and its applications. The constraints on the spectral parameters are tight whenever the foreground signal is high, or the noise level is low or both. The wildly inhomogeneous noise patterns are thus likely to lead to very different constraints depending on the fortuitous overlap of the densely observed sky areas with high foreground areas or lack thereof. Consequently, the predicted levels of both systematic and statistical residuals will not only very wildly but also will depend on the details of modelling, which can be difficult to control. As such they may not be particularly illuminating and useful in the performance forecasting, whatever is the specific way of obtaining those. Indeed, though such coincidental alignments may happen in the analysis of an actual data sets, it is likely that if the constraints on spectral parameters are found to be dominated by a handful of pixels, in which the foreground levels happen to be high and the instrumental noise -- low, the best way forward could be to isolate these pixels in the analysis by for instance assigning to them a new set of parameters.
Obviously, this effect is also present when the noise is homogeneous, however the noise fluctuations  tend to amplify its role. In the former case, the effect is thought to be minimized by usually an implicit assumption that the foregrounds in the observed sky area are typical and that the area is large enough to include a range of typical foreground features. 

The observational strategies of the CMB experiments, including those of the satellite missions~\cite{PlanckMission2011, WmapMission2003}, however, commonly lead to inhomogeneous distributions of the observations over the covered sky area. Therefore, estimating the effects of the inhomogeneities on the forecasts is  of clear importance. One way of proceeding here could be to derive the constraints on spectral parameters not tied up to specific foreground templates but averaging over plausible foreground morphologies. Though, as mentioned earlier this is something what can be readily incorporated in the proposed formalism, we leave an exploration of this aspect of the approach for future work. 

Instead, for time being we make an implicit assumption that both the noise inhomogeneities and frequency scaling laws change only slowly with the position on the sky and across the observed sky, the noise power spectrum is expected to provide a description of the noise of the CMB map, which is sufficient for the estimation of the cosmological parameters based on this map. We calculate this spectrum as,
\begin{eqnarray}
C_\ell^{\rm noise} = \frac{1}{n_{pix}} \sum_p \sigma^2_{p, \, \rm cmb},
\label{eqn:noiseSpec}
\end{eqnarray}
with the pixel-variance as given by Eq.~\eref{eqn:cmbVar}. So the only effect due to the noise inhomogeneity or spatial dependence of the scaling laws, which is taken into account here, is an increased noise level.

In the case of the homogeneous noise and a single scaling law assumed for the entire observed patch, as used in the specific examples studied later on, this can be rewritten as,
\begin{eqnarray}
C_\ell^{\rm noise} = \left[ \left( \mathbf{A}^T \mathbf{N}^{-1}_\ell \mathbf{A} \right)^{-1} \right]_{\hbox{\scriptsize{{\sc cmb} x {\sc cmb}}}}
\label{eq:clNoiseWhite}
\end{eqnarray}
where matrices $N_\ell$ are the harmonic space counterparts of $N_p$ and describe the noise spectra of single frequency maps accounting for their resolution,
\begin{eqnarray}
	\mathbf{N}^{ij}_\ell \equiv \left(w_i\right)^{-1}\,\exp\left( \ell(\ell+1)\frac{\rm FWHM_i^{\,2}}{8\log 2} \right)\delta_i^{\;j}
	\label{eq:white_noise_angular_power_spectra}
\end{eqnarray}
with $\left(w_i\right)^{-1}$ the sensitivity of the frequency channel $i$ in $\left(\mu {\rm K_{\rm RJ}-rad}\right)^2$. We note that the beam effects appearing explicitly in this equation are included in there by hand as they are not fully consistent with the formalism presented earlier, which assumes no pixel-domain correlations. Indeed, the latter requirement implies that no beam-deconvolution procedure of any sort can be applied to the single-frequency maps prior to the component separation and instead the latter would have to be performed instead on the maps smoothed to the largest experimental beam. However, once the spectral parameter estimation is done, and thus the constraints on them are set using only the underpixelized data, both the noise and the residuals can be extrapolated to higher multipoles as long as the number of frequency channels with a resolution high enough is sufficient to ensure that the matrix on the rhs of Eq.~\eref{eq:clNoiseWhite} is invertible. 

\subsection{Parameter estimation}

\label{ssect:parestim}

As before, to estimate the parameters, we proceed as we would have done it if we did not know that our maps are potentially systematically biased.
We therefore start off from the standard Gaussian likelihood, which accounts only for the presence of the noise, the CMB signal and the statistical foreground residual
 in the recovered CMB map, assuming that they all are
Gaussian with the total covariance given by $\mathbf{C}$, i.e.,
\begin{eqnarray}
{\cal S}^{par} \equiv -2\,\ln {\cal L}_{par} = \mathbf{a}^t \mathbf{C}^{-1} \mathbf{a} + \ln \det \mathbf{C}.
\label{eqn:parLike}
\end{eqnarray}
Here, $\mathbf{a}$ is a harmonic representation of the available CMB map. In our case this is the map obtained from the component separation procedure, and which therefore may include
in addition to the CMB signal, the measurement noise and the statistical
residual also systematic bias.
This latter is however ignored in the {\it assumed} data covariance matrix, $\mathbf{C}$, which as in~\cite{Errard2011} includes only first three of these contributions: the CMB signal, the noise and the statistical foreground
residual. 
The explicit form of $\mathbf{C}$ is given later on.
The cosmological parameters we are after here enter only in the expression
for the CMB covariance.
The parameter likelihood averaged over the instrumental noise and CMB signal realizations is given by,
\begin{eqnarray}
\langle {\cal S}^{par}\rangle = {\rm tr} \, \mathbf{C}^{-1} \mathbf{E} + \ln \det \mathbf{C},
\label{eqn:parLikeAv}
\end{eqnarray}
where $\mathbf{E} \equiv \langle \mathbf{a}\mathbf{a}^t\rangle$ is the correlation matrix of the data. 
The values of the cosmological parameters, which maximize Eq.~\eref{eqn:parLikeAv}, are those, for which the likelihood gradient vanishes.
This reads, e.g.,~\cite{Tegmark1997},
\begin{eqnarray}
\langle {\cal S}^{par}_{,i}\rangle  =  {\rm tr} \l[ \mathbf{C}^{-1} \mathbf{C}_{,i} - \mathbf{C}^{-1} \mathbf{C}_{,i} \mathbf{C}^{-1} \mathbf{E} \r].
\label{eqn:parGrad0}
\end{eqnarray}
Given that 
the harmonic coefficients of the CMB map, $\mathbf{a}$, can be represented as,
\begin{eqnarray}
\mathbf{a}_j & = &  \mathbf{a}^{\rm cmb}_j + \mathbf{a}^{\rm noise}_j + \mathbf{\tilde r}^{\rm cmb}_j  \nonumber \\
& = &  \mathbf{a}^{\rm cmb}_j + \mathbf{a}^{\rm noise}_j + \mathbf{\tilde y}_j + \mathbf{\tilde Y}^{\left(1\right)}_{j} {\boldsymbol \delta}
+ {\boldsymbol \delta}^t \mathbf{\tilde Y}^{\left(2\right)}_j {\boldsymbol \delta},
\end{eqnarray}
we can write up to the second order in $\boldsymbol \delta$,
\begin{eqnarray}
	\centering
		\mathbf{E}_{jj'} \hskip -5pt & \equiv &  \hskip -3pt\mathbf{D}_{jj'} +
		  \mathbf{\tilde y}_j \mathbf{\tilde y}_{j'}^*  
		  + \mathbf{\tilde y}_j \, {\rm tr} \l[ \mathbf{\tilde Y}^{\left(2\right) *}_{j'} \, {\boldsymbol \Sigma}\r]\nonumber\\
		   & + & \hskip -3pt {\rm tr} \l[ \mathbf{\tilde Y}^{\left(2\right)}_j  {\boldsymbol \Sigma}\r] \mathbf{\tilde y}^*_{j'}  + \mathbf{\tilde Y}^{\left(1\right) t}_j {\boldsymbol \Sigma} \mathbf{\tilde Y}^{\left(1\right) *}_{j'} \nonumber\\
		   & = & \hskip -3pt \mathbf{D}_{jj'} \hskip -2pt+ \mathbf{\tilde y}_j \mathbf{\tilde y}_{j'}^*+ \mathbf{\tilde y}_j  \, \mathbf{\tilde z}_{j'}^*  + \mathbf{\tilde z}_j \mathbf{\tilde y}^*_{j'} 
		   \hskip -2pt +  \mathbf{\tilde Y}^{\left(1\right)}_j \, {\boldsymbol \Sigma} \, \mathbf{\tilde Y}^{\left(1\right), \, \dagger}_{j'}\hskip -3pt.\ \ \ \ \ 
		   \label{eqn:EmatDef}
\end{eqnarray}
where the cross-terms vanish given that $\langle a_{lm}\rangle = 0$ for both the noise and CMB, and $\langle \delta \beta\rangle = \langle a_{lm}\delta \beta\rangle = 0$ by definition.
Here,
\begin{eqnarray}
\mathbf{D}_{jj'} &\equiv& \langle \mathbf{a}^{\rm cmb}_j\mathbf{a}^{\rm cmb,\,\dagger}_{j'} \rangle +  \langle \mathbf{a}^{\rm noise}_j\mathbf{a}^{{\rm noise}, \, \dagger}_{j'} \rangle \nonumber\\
& = & C_{\ell}^{\rm cmb}\delta_{j j'} + C_{\ell}^{\rm noise}\delta_{j j'} \equiv C_\ell^{\phantom{\ell}} \delta_{jj'},
\end{eqnarray}
is the CMB plus noise only covariance, which is assumed hereafter to be diagonal in the harmonic space, corresponding therefore to the assumption of the stationary pixel-domain noise.

We can rewrite the expression for the true data covariance, Eq.~\eref{eqn:EmatDef}, in the matrix form as,
\begin{eqnarray}
\mathbf{E} = \mathbf{D} + \mathbf{\tilde y} \mathbf{\tilde y}^\dagger + \mathbf{\tilde z} \mathbf{\tilde y}^\dagger + \mathbf{\tilde y} \mathbf{\tilde z}^\dagger + \mathbf{\tilde Y}^{\l(1\r)} \mathbf{\Sigma}\, \mathbf{\tilde Y}^{\l(1\r) \dagger}.
\label{eqn:EdefMat}
\end{eqnarray}
In contrast, the assumed covariance, $\mathbf{C}$,  will be like above but with the terms due to the bias omitted, i.e.,
\begin{eqnarray}
\mathbf{C} = \mathbf{D} + \mathbf{\tilde Y}^{\l(1\r)} \mathbf{\Sigma}\, \mathbf{\tilde Y}^{\l(1\r) \dagger}.
\label{eqn:cMatDef}
\end{eqnarray}

This last expression resembles the one derived in~\cite{Errard2011}, Eq.~(B5), however it extends it by accounting correctly for the possible presence of
multiple foreground components.  
We can now use Eqs.~\eref{eqn:cMatDef} and~\eref{eqn:EmatDef} to calculate the ensemble averages of derivatives of the likelihood given by Eq.~\eref{eqn:parLike}. The full
expressions are quite lengthy and are collected in Appendix~\ref{app:parLikeDervs}. 

We note that though a computation of the explicit form of the covariance matrices, $\mathbf{C}$ and $\mathbf{E}$, 
Eqs.~(\ref{eqn:cMatDef}) and~(\ref{eqn:EdefMat}), respectively,
requires knowledge of all the harmonic modes of the foreground components, for the calculation of the ensemble averaged likelihood and its derivatives, we need only various cross-spectra
of pixel-domain objects defined by $\mathbf{\tilde y}$, $\mathbf{\tilde z}$ and columns of $\mathbf{\tilde Y}^{\l(1\r)}$. This is a general observation, which stems  merely from the assumptions 
about the diagonality of the CMB signal and noise covariance matrices in the harmonic domain, and thus their stationarity in the pixel domain, and does not involve any specific assumptions 
about the foregrounds themselves. This can be intuitively understood, as whenever the CMB signal and noise are both stationary in the pixel domain the constraints on the cosmological parameters 
parametrizing the CMB spectrum can only depend on foreground properties averaged over the observed patch such as their power spectra, and not
on their morphology or phase-dependent information. This is the case, whatever is the actual statistics of the foreground templates.  If the noise is inhomogeneous and anisotropic in the 
pixel domain, it will give preference to some selected modes over others and 
the results of the parameter estimation will depend on both the power and morphology of the foregrounds, making the forecasting dependent on subtle details of the modelling, many of which are still poorly known at this time.
This is similar to the case discussed earlier in the context of the component separation.
Unlike in that latter case now the assumption of the pixel-domain noise stationarity not only makes our forecasts less detail dependent  but it is in fact necessary 
in order to facilitate the analytic calculations, which in turn are essential for the numerical efficiency of the proposed approach. 

Hereafter, we will thus employ the noise spectra as given in Eq.~\eref{eqn:noiseSpec} accepting that some of the information is lost in this process. Our forecast will therefore be pessimistic in some sense but more reliable. The information is not lost only when the noise of the recovered CMB map is homogeneous, which is equivalent to the case of the homogeneous noise in the frequency maps and global scalings laws, and where our assumptions are automatically fulfilled.

We note that more general noise power spectra than the white noise cases can be studied using this formalism, for instance, the spectra with excess power at low-$\ell$ end of the spectrum, devised to mimic the potential effects of time-domain noise correlations and/or time-domain filtering. As no noise correlations are included in our component separation step this will not be a fully consistent approach, however, as the pixel-stationary noise correlations are expected to have a bigger impact on the cosmological parameter estimation than the component separation step, such ad hoc adjustments can be expected to provide useful and meaningful insights.

This part of our algorithm now proceeds as follows: for the given foreground models, true and assumed, and the noise power spectrum, we maximize Eq.~\eref{eqn:parLikeAv} using its first derivatives, Eq.~\eref{eqn:parGrad0},
to find the maximum likelihood-like values of the cosmological parameters, which can be however biased by the presence of the residual foregrounds.
We then use the ensemble average Hessian of the likelihood in Eq.~(\ref{eqn:parLikeAv}), with the true data matrix set to $\mathbf{E}$, to assign uncertainties to these estimates.

The proposed procedure is well-defined in full- or nearly full- sky coverage cases. If only a limited sky area is available, the procedure can be adapted  to produce some meaningful estimates. This involves the usual steps of introducing a low-$\ell$ cut-off corresponding to the largest mode, which can be still well-constrained by the cut-sky data, and of multiplying the derived, full-sky Hessian by the observed sky fraction, $f_{sky}$, to reflect the overall loss of the independent modes in the available data. Were we projecting out all the sky modes potentially contaminated by the foreground residuals, what would correspond to $\boldsymbol{\Sigma} \rightarrow \infty$, this later step would suffice. However, in our case $\boldsymbol{\Sigma}$ is finite and in fact hoped to be small so the spectral parameters are well-determined. Moreover, as it is estimated on the component separation step, it already incorporates the information about the observed sky fraction, as roughly $\boldsymbol{\Sigma} \propto 1/n_{pix} \propto 1/f_{sky}$. To account on that, as an input to the cosmological parameter estimation procedure,
 we use the rescaled matrix of spectral parameter errors, $\boldsymbol{\Sigma'} \equiv f_{sky} \boldsymbol{\Sigma}$. This rescaled error matrix roughly reflects the full sky errors and we use it in our algorithm to compute the bias and the Hessian for this case as described earlier. Once this is done we then rescale the Hessian by $f_{sky}$, so the statistical errors on the cosmological parameters are amplified by $\sqrt{f_{sky}}$.

\subsection{Algorithm}

\label{sect:algo}

\begin{figure}
	\centering
		\includegraphics[width=\columnwidth]{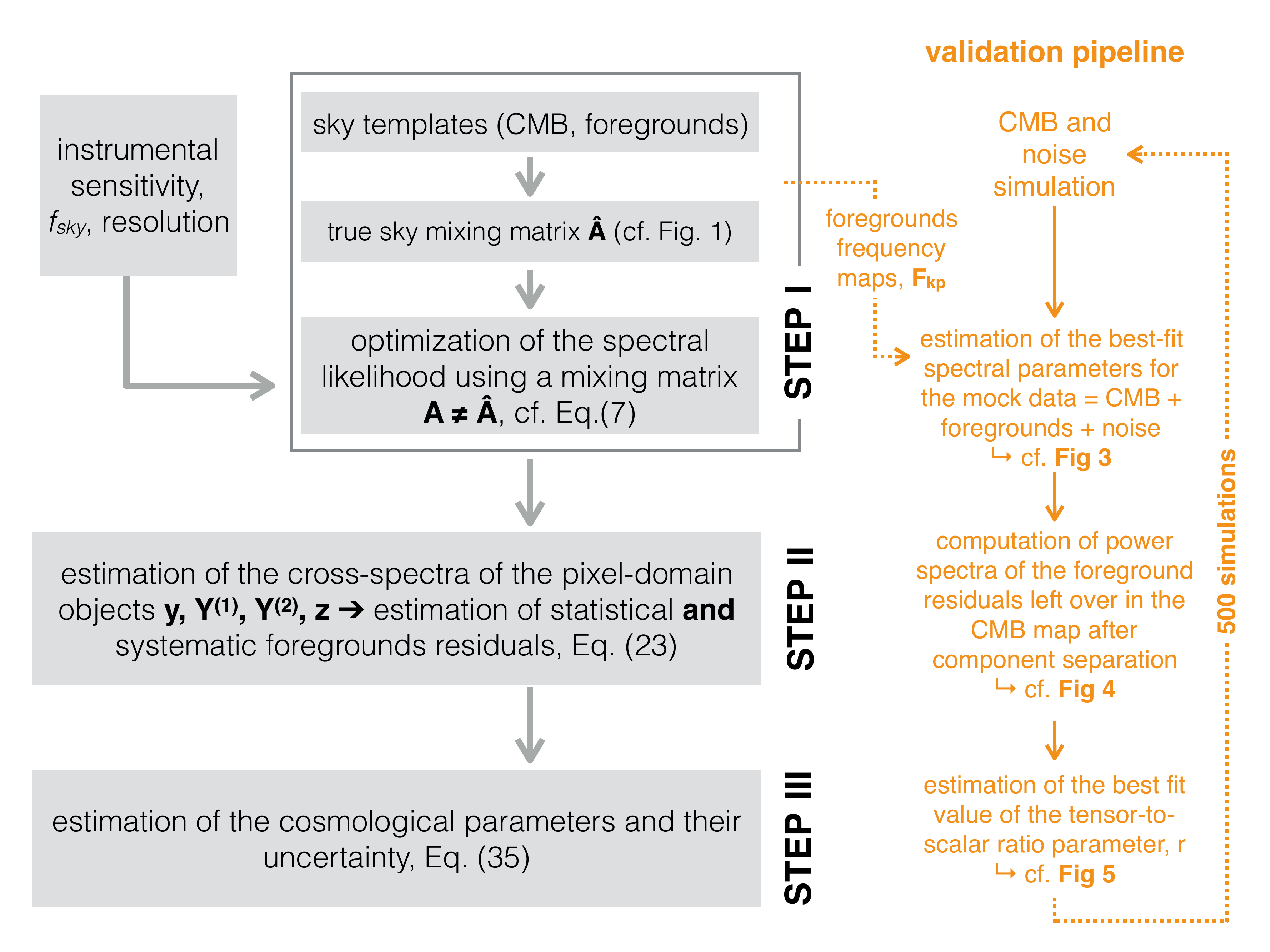}
	\caption{A flow chart of the main steps of the proposed approach and of its validation pipeline implemented in this work.}
	\label{fig:algorithm_schematic}
\end{figure}

\begin{table*}[htb!]
\caption{Instruments specifications}
\centering
\label{table:instrumental_specifications}
\resizebox{12cm}{!}{
\hspace{-2cm}
\begin{tabular}{| l c | c r c r c r c r c r c r c r c r c r c r c r c r c r c r c r |} 
\hline
\hline
frequency [GHz]   & \  & \ &  \hfill \hfill \hfill 40   && \hfill \hfill \hfill 50  && \hfill \hfill \hfill 60 && \hfill \hfill  \hfill 70 && \hfill \hfill \hfill 80 && \hfill \hfill \hfill 90  && \hfill \hfill \hfill 100 && \hfill \hfill \hfill 120 && \hfill \hfill \hfill140 && \hfill \hfill \hfill 165 && \hfill \hfill \hfill 200 && \hfill \hfill \hfill 235 && \hfill \hfill \hfill 280 && \hfill \hfill \hfill 340 && \hfill \hfill \hfill 400 \\
sensitivity [\ukarc] & \ & \  & 42  & &  26 & & 20 && 15 && 12 && 19  &&   12  &&  10   &&    7  &&     7 &&    5  &&    6 && 19  && 10 && 19 \\
FWHM [arcmin]     & \  &\   &108 & & 86 &&  72 && 63 && 55 && 49 && 43  &&     36 && 31  &&   26  &&  22 &&   18 &&    37 && 31  && 26 \\
\hline
\hline
\end{tabular}}
\end{table*}

The approach proposed here involves three main steps,
\begin{description}
\item{\sc step i:} estimation of the spectral parameters, used to parametrize the frequency scaling laws of the components and their uncertainty. 
\item{\sc step ii:} estimation of the cross-spectra of the pixel-domain objects, $\bd{y}$, $\mathbf{Y}^{\l(1\r)}$, $\mathbf{Y}^{\l(2\r)}$, $\mathbf{z}$, characterizing the systematic and statistical residuals.
\item{\sc step iii:}  estimation of the cosmological parameters and their uncertainty.
\end{description}
As the products computed on each of these steps provide the inputs for the next ones, the necessary initial inputs consist of those required for the first step computations. 
In general these are,
\begin{enumerate}
\item the multi-frequency, noiseless true sky signal maps, $\bd{\hat d}$, split into its CMB, $\mathbf{\hat s}$, and foreground, $\mathbf{\hat f}$, parts;
\item the noise covariance matrices for the frequency channels, $\bd{N}$;
\item assumed, parametrizable scaling laws for all considered sky components, $\bd{A}$.
\end{enumerate}
These are in principle sufficient to perform all the steps of the proposed approach. If available, the algorithm proceeds as follows,
\begin{description}
\item{\sc step i:}  The best-fit spectra parameters are found by a direct maximization of the ensemble average likelihood in Eq.~\eref{eqn:likeAvFinal}. This is implemented using a minimization routine from the {\sc Python}'s {\sc Scipy} library implementing the truncated Newton constrained (TNC) solver. It capitalizes on the analytic derivatives of the likelihood with respect to the spectral parameters given by Eq.~(\ref{eqn:profileLikeDer}). The statistical uncertainty is then computed using Eq.~(\ref{eqn:profileLikeSecondDer}).
\item{\sc step ii:} Given the best-fit values of the spectral parameters and their statistical uncertainty estimated on {\sc step i}, we first estimate the pixel-domain objects, $\bd{y}$, $\mathbf{Y}^{\l(1\r)}$, $\mathbf{Y}^{\l(2\r)}$, $\mathbf{z}$, using Eqs.~(\ref{eqn:yVectsDef}) and~(\ref{eqn:zFullPix}), and calculate their spherical harmonic decomposition and cross-spectra as needed;
\item{\sc step iii:} The forecasted values of the cosmological parameters are then computed by directly maximizing the likelihood in Eq.~(\ref{eqn:parLikeAv}) with the first derivative computed with help of Eqs.~(\ref{eqn:gradInit}), (\ref{eqn:gradTerm00}) and (\ref{eqn:gradTerm2}). The likelihood itself is calculated using Eqs.~\eref{eqn:parLikeChi2term}-\eref{eqn:parLikeDetTerm},
These computations use the cross-spectra computed on {\sc step ii} and is performed, as before, with help of a \textsc{Scipy} minimization routine implementing the truncated Newton constrained (TNC) solver. However,  this is now only performed after a rough grid-based search needed to ensure a reasonable starting point. The statistical uncertainty is then computed numerically as the curvature of the ensemble average cosmological parameter likelihood, which is an output of the routine.\\
Alternately, whenever the number of the sought-after cosmological parameters is very limited, typically $\simlt 2$, the proposed formalism permits a full investigation of the likelihood function by a direct evaluation of Eq.~\eref{eqn:parLikeAv} on a grid of the parameters and using the analytic results from Appendix~\ref{app:parLikeDervs}, Eqs.~\eref{eqn:parLikeChi2term}-\eref{eqn:parLikeDetTerm}.
\end{description}
We note that in many cases of interest not all these calculations have to be actually performed and instead can be supplemented by some additional or alternative inputs. For instance, if neither the frequency scaling laws for the assumed sky model nor the pixel-domain noise depend on the pixel position on the sky, the sufficient information about the foregrounds can be provided by the foreground component-component covariance matrix and the component-component cross-spectra, Appendix~\ref{app:simpCase}. In such a case, the cross-spectra of $\bd{y}$, $\mathbf{Y}^{\l(1\r)}$, $\mathbf{Y}^{\l(2\r)}$, $\mathbf{z}$  are directly related to those, simplifying and accelerating the calculations on required on {\sc step ii}. Also, the estimation of the spectral parameters, {\sc step i}, can be then performed using only the component-component covariance matrix of the foregrounds as the input.

A simplified flowchart of the method is shown in Fig.~\ref{fig:algorithm_schematic}.

\section{Validation and demonstration}

\label{sect:applic}

We validate our approach using simulated, multi-frequency data sets of a putative CMB observation. Below we describe in turn: the assumed experimental set-up, the assumed foreground models, and the adopted validation methodology.

\subsection{Set-up}

\subsubsection{Observation}
We assume here a nearly full-sky observation, which have produced a set of multi-frequency maps of $70$\% of the entire sky suitable for cosmological component separation. The frequency bands, their assumed resolutions and sensitivites are all listed in Table~\ref{table:instrumental_specifications} and have been selected following loosely the example of a contemporary satellite mission concept,~\footnote{{\tt http://ltd16.grenoble.cnrs.fr/IMG/UserFiles/Images/\ 09\_TMatsumura\_20150720\_LTD\_v18.pdf}}. Top-hat band passes are used throughout with their widths set to $30$\% of the central frequency for each band. We assume the duration of the observation to be 3 years, the noise to be white in map domain, with the homogeneous sky coverage, and no correlation between the maps at different frequencies included. In the following analyses we always include harmonic modes with $\ell$ ranging from $2$ to $500$, with the high-end cut-off set by the assumed resolutions.
We emphasize that our goal here is to demonstrate and validate the proposed method and not to provide a performance evaluation of any specific instrument, what is left to the future work.

The noise spectrum, $C^{\rm noise}_\ell$, characterizing the noise of the final recovered CMB map, is  computed using Eqs.~\eref{eq:clNoiseWhite}~and~\eref{eq:white_noise_angular_power_spectra} with the parameters listed in Table~\ref{table:instrumental_specifications}.

\subsubsection{Sky models}

We adopt, as the true sky model, a model composed of two diffuse foreground components, one dust-like and the other synchrotron-like. We represent them as templates at $150$GHz and scale them following the scalings laws as described below to all the frequencies of interest. For the templates we used the so-called COMMANDER dust and synchrotron maps, scaled to 150GHz using the Planck's fiducial scaling laws, as included and described in the Planck's latest polarized data release~\footnote{\protect\url{http://pla.esac.esa.int/pla}}.  For simplicity, but also in agreement with recent findings of~\cite{2016arXiv160607335P}, hereafter we restrict ourselves to the case with no spatial variability of the scaling laws. We consider only $70\%$ of the sky leaving out the Galactic plane as well as some other high foreground regions as defined by the sky mask provided by the Planck collaboration~\footnote{\protect\url{http://pla.esac.esa.int/pla}}.
 
\label{ssect:appMismatch}

\begin{figure}
	\centering
		\includegraphics[width=1.15\columnwidth]{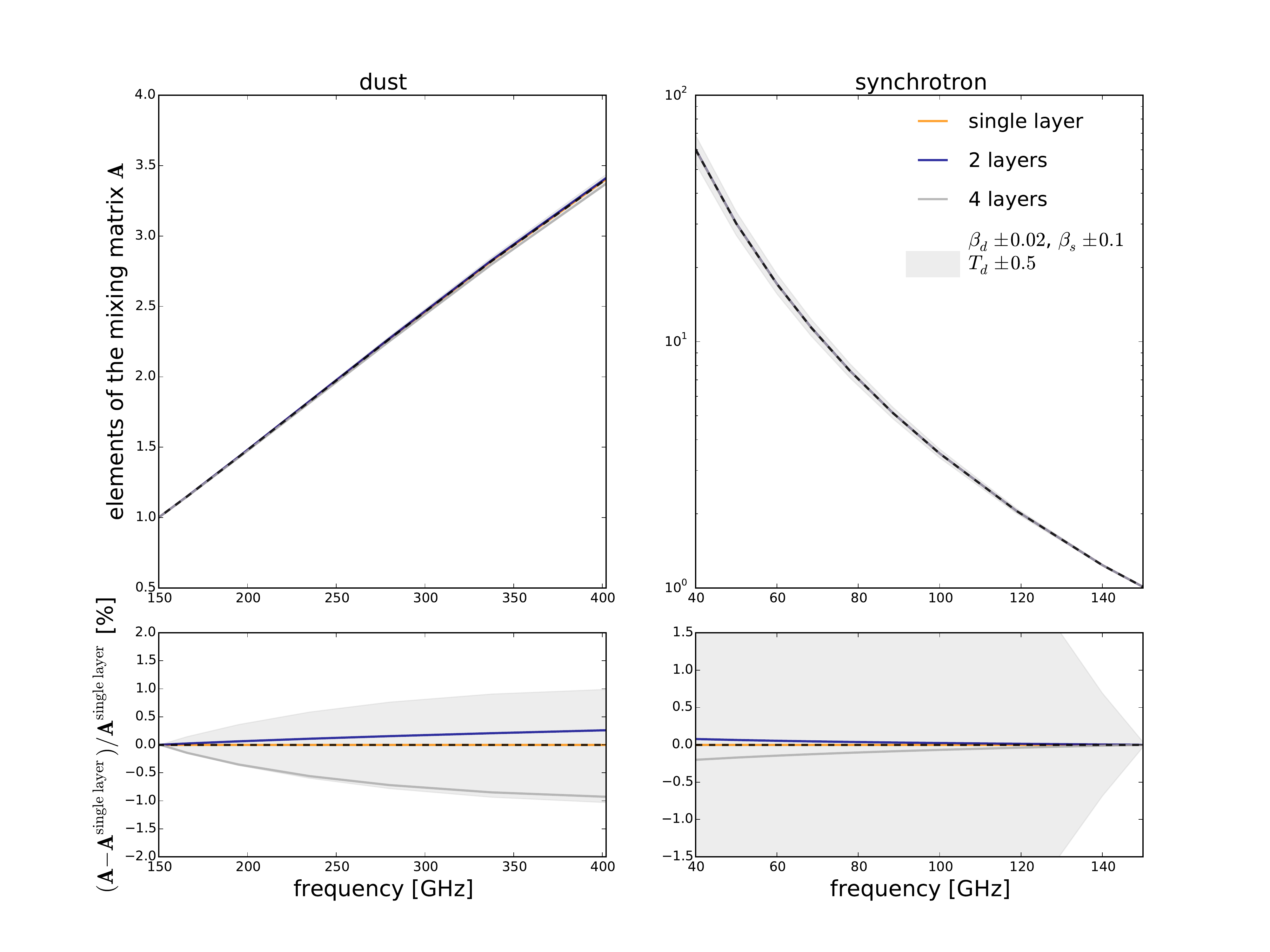}	
	\caption{Frequency scaling laws used in this work to demonstrate our method. The top panels show the scaling laws for dust, left, and synchrotron, right, for three different cases involving mixtures of 1, 2 and 4, grey-bodies, for dust, and power-laws, for synchrotron.  The bottom panels show the same lines but relatively to the single term case. Shaded areas show a rough, 1-$\sigma$, uncertainty on these scaling laws consistent with the  Planck data~\cite{2015AA...576A.107P}.}
	\label{fig:scalings}
\end{figure}

We model dust frequency scaling law as a sum of multiple grey-body terms each computed with a different set of grey-body parameters: the power law index, $\beta_d$, and temperature, $T_d$. Similarly, we model synchrotron frequency scaling as a sum of power laws with different power law indices, $\beta_s$. Though these models are very simple, they allow us to investigate a range of different cases from the simple scaling law models, involving only a single term, to progressively more complex ones based on two and more terms. Specifically, in the following we will focus on three cases of the true sky scaling laws: one involving only one term, with the parameters set to be $T_d = 19.6$K, $\beta_d = 1.59$, and $\beta_s = -3.1$, and two more complex scalings based on a combination of 2 and 4 terms respectively for both dust and synchrotron. The effective scalings are shown in Fig.~\ref{fig:scalings}, where the top panels show the absolute scaling laws and the bottom ones -- these laws relative to the single-term case. Clearly, the two-term scaling leads to the departures of up to $\sim 0.5$\% in the dust scaling within the consider range of frequencies, while the 4-term cases admits deviations as big as $\sim 1$\% for both the dust and synchrotron. The specific parameters used for the calculations of all the terms have been tuned to ensure that the adopted scaling laws are consistent with the Planck constraints~\cite{2015AA...576A.107P}, but also to allow us to validate and demonstrate our method in qualitatively different regimes. We discuss this in more detail in the next Section.

In contrast, while performing the component separation step on the simulated data sets, we always assume single-term-only scalings for both dust and synchrotron, fixing the dust temperature, $T_d$, to the actual value for the first term, i.e., $T_d = 19.6$K, and let the data determine values of two spectral indices, $\beta_d$ and $\beta_s$.

\subsection{Validation procedure}

\label{ssect:validProc}

We validate our approach using simulations.
These assume the foreground model described above.
Each foreground template is scaled to each considered frequency band using one of the three scaling laws and integrated over the frequency band passes. These are coadded with $500$ simulated CMB maps produced as a random realization of the Gaussian process with power spectra defined by the standard cosmological model with parameters set to the best fit Planck values.  All the signal maps are downgraded to $N_{\rm side}=32$. We thus obtain three sets of $500$ simulated, single-frequency maps, each set implementing a different foreground scaling law.  In addition, for each set we generate $500$ independent realizations of the instrumental noise, which combined together with the signal maps create mock data sets.  These noise realizations are drawn for each frequency band separately and modelled as  a Gaussian process with variance as given in Table~\ref{table:instrumental_specifications} and zero mean.  There are no other systematic effects included in the simulations.

\begin{figure}
	\centering
		\includegraphics[width=1.0\columnwidth]{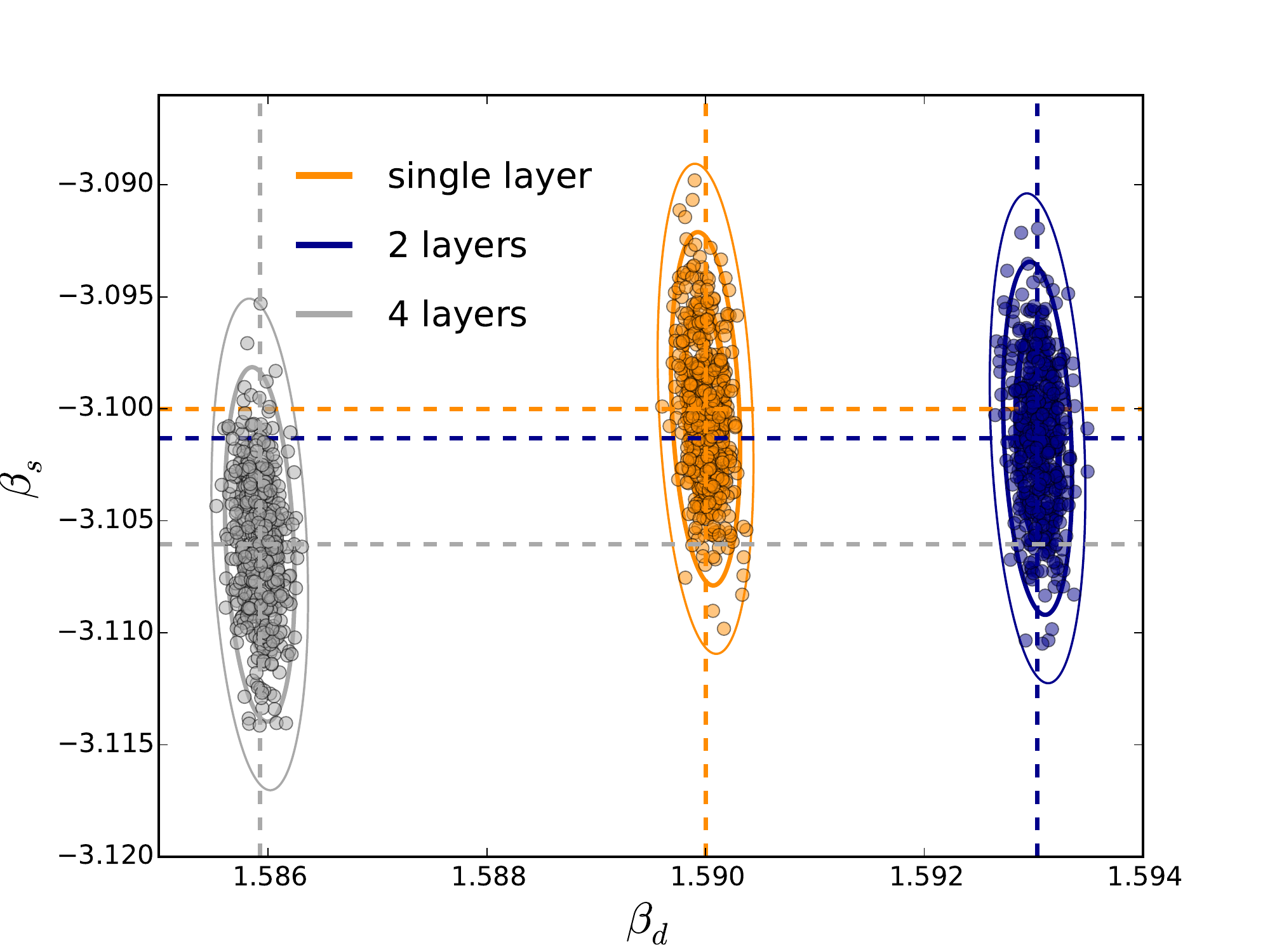}
	\caption{Constraints on spectral parameters, power-law indices for dust, $\beta_d$, and synchrotron, $\beta_s$, forecasted for the assumed multi-frequency observation using the approach presented here. These are shown as solid ovals corresponding to $1, 2,$ and $3$- $\sigma$ contours of the spectral parameter likelihood and computed for the three different foreground models as discussed in the text and shown in Fig.~\ref{fig:scalings}. Each model results are shown in different colors as indicated in the legend. The filled circles show the results of a direct maximization of the spectral likelihood performed for $500$ independent realizations of the considered data set. The thin dashed lines show the position of the likelihood peaks as determined by the semi-analytic approach. These are virtually indistinguishable from the average values derived from simulations.}
	\label{fig:spectralParams}
\end{figure}

As part of the validation procedure, we analyze each of the simulations as we would the actual data, performing first the pixel-domain, maximum-likelihood, parametric component separation followed by the pixel-domain, maximum-likelihood cosmological parameter fitting.  At the conclusion of each of these two steps we compare the simulated results with the corresponding results obtained with the proposed method. 
Specific comparisons, performed on each step of the proposed algorithm, are as follows, cf. Fig.~\ref{fig:algorithm_schematic}: 
\begin{description}
\item{\sc step i:} 
we estimate the best-fit spectral parameters for the mock data by explicitly maximizing the spectral likelihood in Eq.~(\ref{eqn:profileLikeDef}). We do so for each of the three sets of the $500$ simulations. We then compare the results with the expected distribution of the measured spectral parameters derived semi-analytically using Eqs.~(\ref{eqn:likeAvFinal}) and~(\ref{eqn:specParError}) and assuming a Gaussian approximation.

\item{\sc step ii:} 
to validate this step we compute power spectra of the foreground residuals left over in the CMB maps after the component separation step. We note that these are higher level objects, which are neither explicitly derived on this step of the processing nor needed for the subsequent stages of the procedure. However, they  combine the same information as the direct products and have well-defined physical interpretation. They therefore provide a meaningful and intuitive comparison metric. \\
We implement this comparison as follows.
For the simulated data, we first compute the cleaned CMB map using expression in Eq.~(\ref{eqn:compSepEq}) assuming the spectral parameters as derived on {\sc step i}. The map domain residual is then derived by subtracting from this map the true CMB map used to simulate the input data. We then calculate the power spectrum of the residuals and compare it with the semi-analytic results derived via Eqs.~(\ref{eqn:psResFinal}) and~(\ref{eqn:varResFinal}). 

\item{\sc step iii:} we estimate the best fit value of the tensor-to-scalar ratio parameter, $r$, by explicitly maximizing full cosmological parameter likelihood, Eq.~(\ref{eqn:parLike}), we bin the results and compare the histogram with the Gaussian distribution with an average and dispersion derived using the proposed approach, Section~\ref{ssect:parestim}.
In addition, we explicitly compute the averaged likelihood as a function of $r$, using Eq.~\eref{eqn:parLikeAv}. In both these latter cases the computations are performed directly in the harmonic domain using the harmonic space representation of the cleaned CMB maps obtained on {\sc step ii}. This is done with help of a standard spherical harmonic transform and thus neglects the effects due to cut sky.  This is expected to lead to some power loss at the low-$\ell$ end and thus may affect the level of bias in the estimated values of $r$. This should be taken care of in the actual forecasting process however is irrelevant for the formalism demonstration purposes.

\end{description}

Due to computational-time limitation we perform the end-to-end analysis of the actual simulation on underpixelized maps with {\sc HEALpix} $n_{side} = 32$. This is what restricts the analysis presented hereafter to $\ell_{max} = 64$ as the semi-analytic approach can be easily applied for much higher cut-offs.

\subsection{Results}

\label{ssect:appResults}

\begin{figure*}
	\centering
		\includegraphics[width=2\columnwidth]{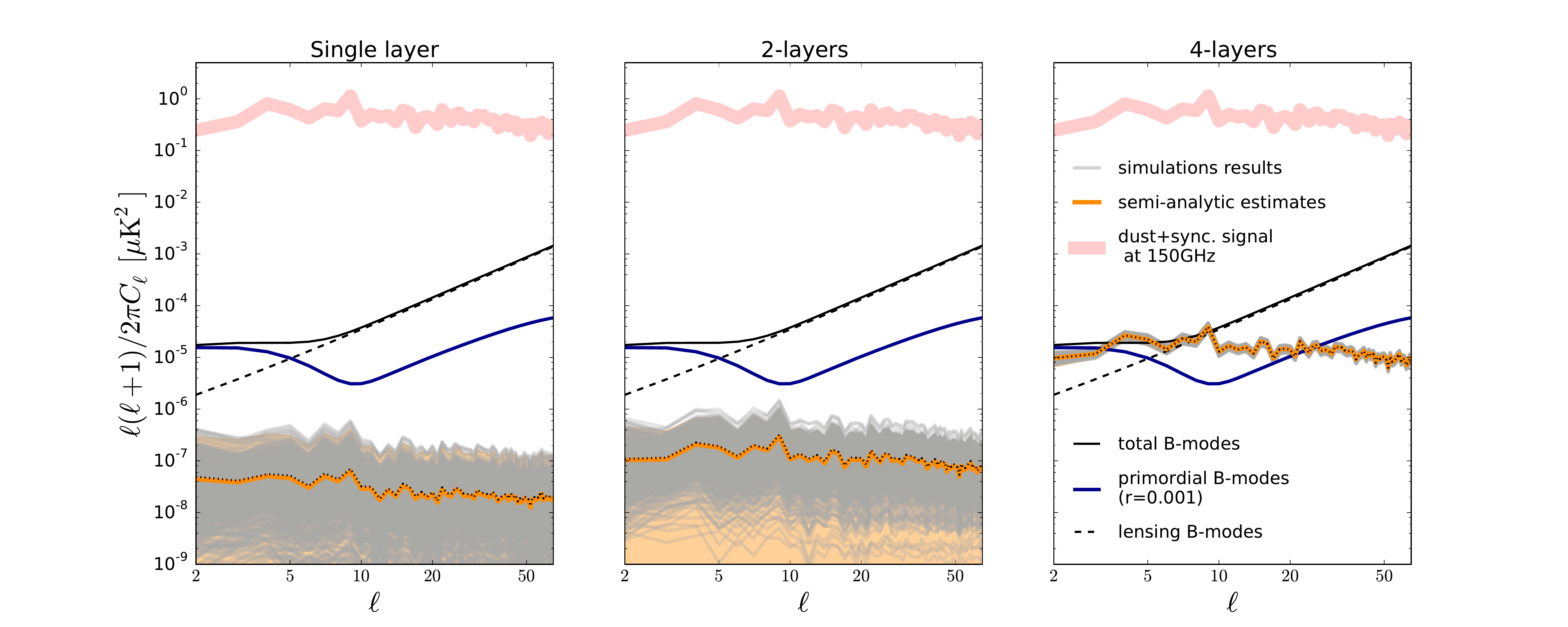}
	\caption{Foreground residuals present in the cleaned CMB maps after the parametric component separation. The panels corresponds to the three different scaling laws considered in this work. The results shown in orange are obtained using the semi-analytic method proposed in this work with the solid orange lines showing the ensemble-averaged residuals and the shaded orange areas depicting the $2\sigma$ scatter. The grey lines show the residuals computed case-by-case for each of the simulations and the thick, dashed lines show their averages. The theoretical $B$-mode spectra showing the primordial ($r=10^{-3}$), dark blue line, lensing, dashed black, and total, thick black line, are also shown for reference. The three models represent cases with dominant statistical residuals, left panel, dominant systematic residuals, right panel, and comparable statistical and systematic residuals, middle panel. For comparison the red-shaded band shows the total, dust plus synchrotron, foreground  signal at $150$GHz.
	Only the low-$\ell$ residuals are shown here.}
	\label{fig:spectraRes}
\end{figure*}

\begin{figure*}
	\centering
		\includegraphics[width=2\columnwidth]{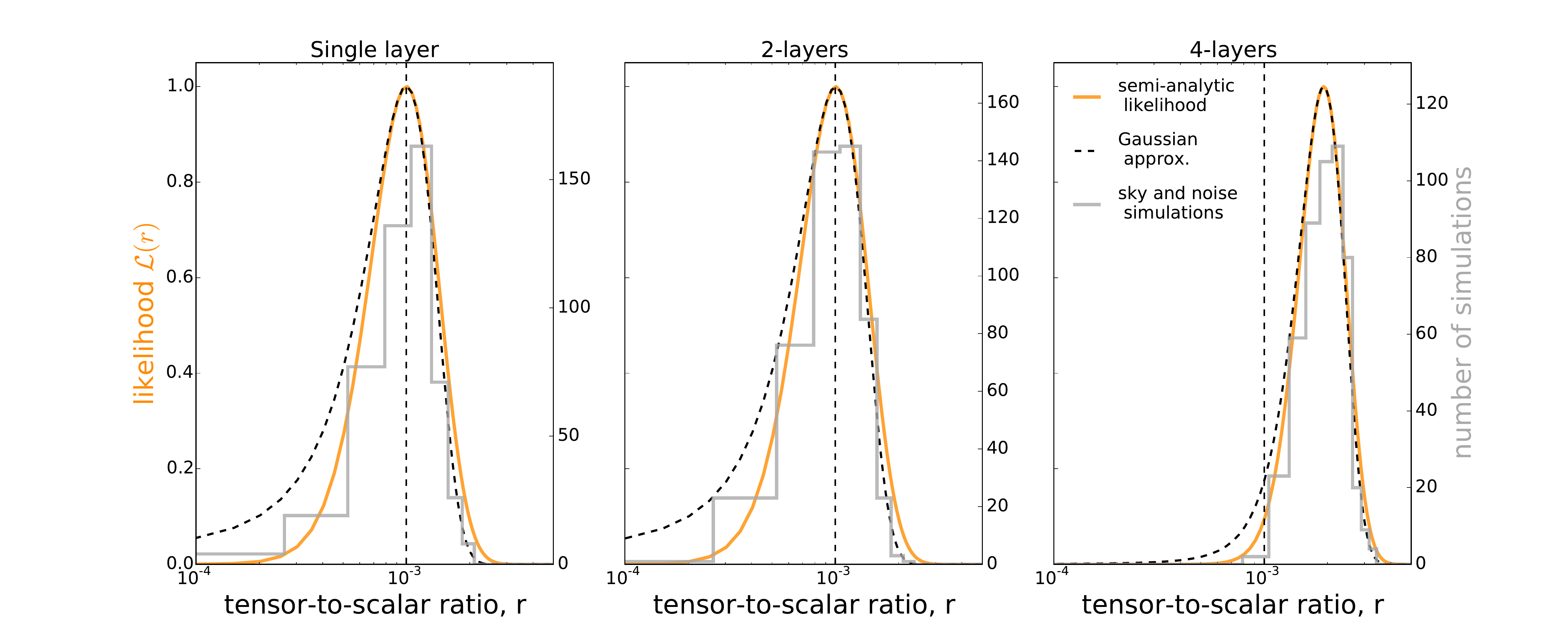}		
	\caption{Forecasted constraints on the tensor-to-scalar ratio parameter, $r$, derived from foreground-cleaned CMB maps derived for the multi-frequency observation studied here. The orange solid lines show the likelihoods on  $r$ averaged over the statistical ensemble of noise and CMB realizations estimated using Eq.~\eref{eqn:parLikeAv}. The black, thick, dashed lines show the results of the Gaussian approximation with the average and variance computed semi-analytically as described in this work. The histograms show results of an end-to-end analysis of the putative, simulated data sets involving random realizations of the instrumental noise and CMB signal and the foreground contributions as used in the semi-analytic approach. The results in the left panel show no bias in the estimated value of $r$. The bias in the middle panel though formally non-zero is negligible as compared to the statistical error. In contrast, the bias seen in the rightmost panel is significant as compared to the true value of $r$ and the estimated statistical uncertainty. These results demonstrate an excellent agreement between our semi-analytic approach and the full computation with the Gaussian approximation being, however, more permissive as far as low values of $r$ are concerned. }
	\label{fig:rHistos}
\end{figure*}

We present here results obtained for each of the three steps defined earlier. They are visualized in Figs.~\ref{fig:spectralParams}, ~\ref{fig:spectraRes}, and~\ref{fig:rHistos}.

In Fig.~\ref{fig:spectralParams} we show the spectral laws parameters and their $1,2,3-\sigma$ confidence levels, shown as contours, obtained as a Gaussian approximation to our spectral likelihood, Eq.~(\ref{eqn:likeAvFinal}). This likelihood is averaged over possible noise realizations. The thick points show the results of the simulations and therefore each point shows the values of the spectral parameters derived by a direct maximization of the spectral likelihood, Eq.~\eref{eqn:profileLikeDef}. We note that each set of the mock multifrequency data contains different realization of the noise and the CMB signal, however as discussed earlier on, the latter does not impact the spectral parameter estimation, see  Eq.~(\ref{eqn:resCMBpix}), and the scatter of the spectral parameters values from the simulations seen in the figure is driven only by the noise.
Overall, we see very good agreement between the semi-analytic contours and the overall distribution of the results derived from the simulations, which as expected cover the areas delineated by the contours aggregating around the expected peaks of the likelihood. More quantitively, the average spectral parameters derived from the simulations are the dashed lines and show a good agreement with the semi-analytic values.

We note that the position of the peak of the likelihood is different for each of the cases.
Indeed, only in the case of the single-term model  the true and assumed sky agree and the recovered values of the spectral parameters agree with those used in the sky simulations. For the multi-term scaling laws, the assumed model does not provide an accurate description of the true scaling laws and the derived values of the spectral parameters do not carry any more any physical meaning, but rather they are some effective values that lead to the scaling laws matching the true ones most accurately.

In Fig.~\ref{fig:spectraRes} we show the comparison performed after {\sc step ii}. The grey lines show power spectra of the foreground residuals computed for each of the $500$ realizations of the noise and the thick dashed lines show their average. The solid orange lines depict the average residual spectra and the orange shaded areas show $2\sigma$ uncertainty computed with help of the proposed method.  We see a very good overall agreement for all three choices of the scaling laws.

The residuals shown in the left panel are merely due to the statistical scatter in the estimates of the spectral parameters and this case corresponds to those studied in~\cite{Errard2011}. Though in this case no bias is expected on the map level, this is not so on the power spectrum level. Indeed, the average power spectrum of the residuals does not vanish as it is indeed shown in the figure by both semi-analytic and simulated results.

In the middle and right panels, given the scaling laws mismatch the residuals are due to both the systematic and the statistical errors. In the middle panel, both these errors are comparable and non-negligible. While in the right panel, the systematic contribution is by far dominant. These results show that the systematic residuals due to the mismatch in the scaling laws can quickly dominate over the statistical residuals even for rather minor levels of the mismatch ($0.5$\% and $1$\% in the cases shown in the middle and right panels) at least as long as the freedom of introducing more spectral parameters is not capitalized on. 

The residuals may potentially affect the values of the cosmological parameters determined from the CMB maps cleaned with the parametric component separation approach. This is not the case for the statistical residuals as in the case shown in the left panel of Fig.~\ref{fig:spectraRes}. This is because our parameter likelihood, Eq.~(\ref{eqn:parLike}), is written in the map-domain, where the foreground residual averaged over ensemble of the noise realization vanishes, and because it accounts for the extra statistical uncertainty. In this case we thus expect only the extra uncertainty but no bias for the estimated cosmological parameters. This would not have been the case, were our likelihood written in the power spectrum domain.
The bias of the cosmological parameters is expected in our approach once the systematic foreground residuals are present. In such cases, the cosmological parameter estimates will be affected by both biases and extra uncertainty. 

Fig.~\ref{fig:rHistos} demonstrates all these general considerations in the context of a determination of parameter $r$. In this figure, the smooth orange lines show the predictions obtained from our semi-analytic approach, while the histograms are obtained by performing simplified maximum likelihood parameter fitting applied to the CMB maps contaminated with the foreground residuals. Again we find a very good qualitative and quantitive agreement in all three cases.  The peak value of the likelihoods and histograms shifts progressively away from the true value, taken here to be $r=10^{-3}$, when the scaling law mismatch and therefore the systematic residual is getting bigger. The case shown in the left panel and thus affected only by the statistical residual does not lead to any bias in $r$. In the intermediate case, middle panel, the bias is marginal and negligible when compared to the statistical scatter, however, in the case shown in the right panel the bias is already statistically important. However, even in this case the bias on $r$ is not as large as one may have expected from the level of the residuals in the power spectrum domain as seen in Fig.~\ref{fig:spectraRes}. This merely reflects the fact that power spectra of the foregrounds and CMB signals are sufficiently different that the effects of the former are minimized in the parameter fitting procedure.
Last but not least, we note that the Gaussian approximation tends to underestimate the actual significance of the detection due to the long tail for values of $r$ going to $0$.

With regard to the methodology, in the studied cases we have found that the terms containing either vectors $\mathbf{z}$ or $\mathbf{Y^{\l(2\r)}}$, Eqs.~\eref{eqn:zFullPix} and~\eref{eqn:yVectsDef},  and which thus arise due to our inclusion of  the second order terms in the expansion of the residuals with respect to the spectral parameter deviations, Eq.~\eref{eqn:resTaylor2}, tend to be subdominant and can probably be safely discarded, what would simplify the numerical implementation. On the other hand, neglecting the off-diagonal terms in the covariance matrices, $\mathbf{C}$ and $\mathbf{E}$, Eqs.~\eref{eqn:cMatDef} and~\eref{eqn:EdefMat}, is more consequential as it can potentially cause mis-estimation of the bias in the estimated parameters by as much as $100$\%. At the same time, this will not however lead to a spurious bias if such is absent in the cleaned CMB map.

\section{Conclusions and prospects}

In this work we have proposed a semi-analytic approach suitable for realistic forecasting of constraints, which can be set on the cosmological parameters by multi-frequency CMB experiments in the presence of complex foreground
contaminations. The derived constraints are averaged over the instrumental noise and CMB realizations and consist of the estimates of the most likely values of the parameters as well as of their dispersion.

The method assumes that the foregrounds are cleaned using a pixel-based, parametric, maximum likelihood component separation approach, however it does not require that the parametric model assumed for the separation process matches the true one for any set of parameter values. If the mismatch is indeed present, the estimated scaling laws will differ from the actual sky ones in a systematic way. This leads to foreground residuals, both systematic and statistical, which will be present in the cleaned CMB map. In our approach we first estimate both these residuals and subsequently incorporate them in the pixel-based cosmological parameter likelihood, which we use to set constraints on cosmological parameters. The constraints derived in this way therefore include both biases as well as statistical uncertainty. In this sense our method generalizes previous efforts of the similar kind~\cite{Errard2011, Errard2012, Errard2015}. We have validated the method in the case of pixel-independent scaling laws and white pixel-domain noise, however, the presented algebraic framework is flexible enough to allow for spatial variation of the foreground scaling for both the true and modelled signals as well as some other real life effects. 
Furthermore, we also note that the proposed formalism permits incorporating any uncertainties in the foreground modelling in the final forecasts. This could be a potentially very handy feature if broad families of  the foreground models need to be investigated. We leave detailed studies of those cases for future work.

In the cases studied in this work, we have found that even a rather minor mismatch, say of $\sim 1$\%, between the true and assumed scaling laws over a broad range of frequencies can lead to substantial biases of the estimated value of the tensor-to-ratio parameter, $r$, if its true value is as low as $10^{-3}$. This emphasizes two things: (1) importance of accurate and realistic modelling of the underlying foreground signals in ensuring that the obtained forecasts are realistic; (2) importance of suitably chosen, parametric scaling models. In this work, for the demonstration purposes, we have adopted rather simple models in both these instances. In particular, we have employed a simple, two-parameter scaling model for the separation stage and thus have not explored all the constraining power of the considered observation, which allows for a significantly larger number of the spectral parameters. For these reasons the results shown here should not be seen as a fair evaluation of the performance of the assumed instrumental set-up but rather merely as indicative of more qualitative effects and dependences one may expect in such circumstances.
Again we leave exhaustive explorations of this kind to future work.

Our approach, though clearly more involved and complex than that of~\cite{Errard2011, Errard2012, Errard2015}, retains the speed and efficiency of these previous, simplified techniques, while permitting to attain a higher level of realism. Indeed, all the numerical computations scale linearly with the high-$\ell$ cut-off, $\ell_{max}$, allowing the calculation to be conducted efficiently even for high-resolution experimental set-ups. 
Consequently, the method is very well-suited for optimizations of experimental set-ups and forecasting their performance, in particular whenever large parameter space of  experimental characteristics needs to be considered. 
Equally importantly, this approach also allows for a direct  exploration of a large number of viable foreground models, thus enabling investigations of robustness of the  predictions with respect to details of the foreground modelling -- a key feature given our present ignorance about the polarized foreground emissions in the microwave band and the impact of the assumed foreground models on the derived predictions.
In all these aspects, the proposed approach is complementary to a more thorough but also more resource demanding, fully-fledged, end-to-end analysis of the realistic simulations.
\\
\acknowledgements
We acknowledge use of the {\sc HEALpix}~\cite{gorski2005} package, as well as the \textsc{CMB4cast} code~\footnote{\protect \url{http://portal.nersc.gov/project/mp107/index.html}}.
The work of JE was performed within the Labex ILP (reference ANR-10-LABX-63) part of the Idex SUPER, and received financial state aid managed by the Agence Nationale de la Recherche, as part of the programme Investissements d'avenir under the reference ANR-11-IDEX-0004-02.

\appendix

\begin{widetext}
\section{Spectral likelihood derivatives}

\label{app:mapLike}

We calculate here the first derivative of the likelihood averaged over the
statistical ensemble of noise. Without losing the generality, for the derivation's sake, we assume
only one pixel, and thus drop subscript $p$ and take the noise covariance, $\bd{N}$,
 to be the identity. We then present the fully general expressions only at the very end.

We start from Eq,~(\ref{eqn:profileLikeDef}) and
rewrite it as
\begin{eqnarray}
{\cal S}_{spec}\hskip -2pt 
=
{\rm tr } \l[ \l( \bd{1} - \bd{P} \r)
\bd{d} \bd{d}^t \r].
\label{eqn:profileLikeDefWTrace}
\end{eqnarray} 
Consequently all we need to do is  to compute $\bd{P}_{,\beta}$ and $\bd{P}_{,\beta\beta'}$.
On defining, $\bd{M} \equiv (\bd{A}^t \bd{A})^{-1}$, these can be written as, 
\begin{eqnarray}
  \bd{P}_{,\beta} &=& 
  - \bd{A} \bd{M}\bd{A}_{,\beta}^t \bd{P} +
  {\rm transpose},
  \label{firstDerP}
\end{eqnarray}
and,
\begin{eqnarray}
  \bd{P}_{,\beta\beta'} &=&
  - \bd{P}\bd{A}_{,\beta'} \bd{M}  \bd{A}^t_{,\beta}\bd{P}
  + \bd{A} \bd{M}  \bd{A}_{,\beta'}^t \bd{A} \bd{M}  \bd{A}_{,\beta}^t \bd{P}  
  - \bd{A} \bd{M}  \bd{A}^t_{,\beta\beta'}\bd{P} \nonumber \\
 &+& \bd{A} \bd{M}  \bd{A}^t_{,\beta} \bd{A} \bd{M} \bd{A}_{,\beta'}^t \bd{P}
  + \bd{A} \bd{M}  \bd{A}_{,\beta}^t \bd{P} \bd{A}_{,\beta'} \bd{M} \bd{A}^t
  + {\rm transpose}.
  \label{secondDerP}
\end{eqnarray}
We can now combine all these terms to form the derivatives of the spectral likelihood, obtaining,
\begin{eqnarray}
  \left\langle \frac{\partial {\cal S}_{spec}}{\partial \beta}\right\rangle  \hskip -2pt 
=
\sum_p 
{\rm tr } \l[
  \bd{N}_p^{-1} \bd{A}_p (\bd{A}_p^t \bd{N}_p^{-1} \bd{A}_p)^{-1} \bd{A}_{p,\beta}^t \bd{P}_p
\langle \bd{d}_p \bd{d}_p^t \rangle \r],
\label{eqn:profileLikeDer}
\end{eqnarray} 
and
\begin{eqnarray}
  \left\langle \frac{\partial^2 {\cal S}_{spec}}{\partial \beta \partial \beta'}\right\rangle  \hskip -2pt 
& = &
\sum_p 
{\rm tr }  \Big\{\Big[ \,
 \bd{P}_p\bd{A}_{p,\beta'} (\bd{A}_p^t \bd{N}_p^{-1} \bd{A}_p)^{-1}
  \bd{A}^t_{p,\beta}\bd{P}_p + \bd{N}_p^{-1} \bd{A}_p (\bd{A}_p^t \bd{N}_p^{-1} \bd{A}_p)^{-1}
  \bd{A}^t_{p,\beta\beta'}\bd{P}_p  \nonumber\\
 &- & \bd{N}_p^{-1}\bd{A}_p (\bd{A}_p^t \bd{N}_p^{-1} \bd{A}_p)^{-1}
  \bd{A}_{p,\beta'}^t \bd{N}_p^{-1} \bd{A}_p (\bd{A}_p^t \bd{N}_p^{-1}
  \bd{A}_p)^{-1}  \bd{A}_{p,\beta}^t \bd{P}_p  \nonumber\\  
  &- & \bd{N}_p^{-1} \bd{A}_p (\bd{A}_p^t \bd{N}_p^{-1} \bd{A}_p)^{-1}
 \bd{A}^t_{p,\beta} \bd{N}_p^{-1} \bd{A}_p (\bd{A}_p^t \bd{N}_p^{-1} \bd{A}_p)^{-1}
 \bd{A}_{p,\beta'}^t \bd{P}_p \nonumber\\
 &- &\bd{N}_p^{-1}  \bd{A}_p (\bd{A}_p^t \bd{N}_p^{-1} \bd{A}_p)^{-1}
  \bd{A}_{p,\beta}^t \bd{P}_p \bd{A}_{p,\beta'} (\bd{A}_p^t \bd{N}_p^{-1}
  \bd{A}_p)^{-1} \bd{A}_p^t \bd{N}_p^{-1}\Big] \langle \bd{d}_p \bd{d}_p^t \rangle\, \Big\},
\label{eqn:profileLikeSecondDer}
\end{eqnarray} 
where we have used the fact that the trace of a product of a symmetric matrix and an arbitrary matrix is the same as that of the symmetric matrix and the transpose of the arbitrary matrix. We note that these equations agree with Eqs.~(4) and~(A9) of \cite{Errard2011}. We also note that as in the case studied in this latter work neither derivative depends on the specific CMB sky signal included in the data, $\mathbf{\hat d}_p = \mathbf{\hat A}\mathbf{\hat{s}}$, as long as the CMB frequency scaling is assumed to be known. This can be seen on observing that the sky signal, $\mathbf{\hat s}$, in the expressions for the first and second derivative of the likelihood is processed either by operator, 
\begin{eqnarray}
\bd{A}_{p,\beta} \,\bd{A}_p^t \bd{N}_p^{-1}  \bd{A}_p)^{-1} \bd{A}_p^t \bd{N}_p^{-1} \mathbf{\hat A}_p,
\end{eqnarray}
or the projection operator, $\mathbf{P}_p$.
However, by assumption we have, $\mathbf{A}_{i0} = \mathbf{\hat{A}}_{i0} = 1$,  where subscript $0$ denotes the column corresponding to the CMB and we have adopted the thermodynamical units, and therefore (see also Eq~\eref{eqn:waEqn}),
\begin{eqnarray}
\l[ (\bd{A}_p^t \bd{N}_p^{-1}  \bd{A}_p)^{-1} \bd{A}_p^t \bd{N}_p^{-1} \mathbf{\hat A}_p \r]_{i0} = \delta_{i0}.
\end{eqnarray}
Moreover, given that, $[\bd{A}_{p,\beta}]_{0i} = 0$,
\begin{eqnarray}
\l[ \bd{A}_{p,\beta} (\bd{A}_p^t \bd{N}_p^{-1}  \bd{A}_p)^{-1} \bd{A}_p^t \bd{N}_p^{-1} \mathbf{\hat A} \r]_{0i} = 0
\end{eqnarray}
and the operator removes all the CMB signal present in the input data vector, $\mathbf{\hat{s}}_p$. Similarly, the projection operator, $\mathbf{P_p}$,  projects out the CMB signal in its entirety as,
\begin{eqnarray}
\l[\mathbf{P}_p\mathbf{\hat A}\r]_{i 0} =  \l[\mathbf{N}_p^{-1}\r]_{ij}\,\l[\mathbf{\hat A}_p\, - \,  \bd{A}_p (\bd{A}_p^t \bd{N}_p^{-1} \bd{A}_p)^{-1} \mathbf{A}_p \mathbf{N}^{-1}_p \mathbf{\hat A}_p\r]_{j 0}  \, = \, \l[\mathbf{N}_p^{-1}\r]_{ij} \l(\mathbf{\hat{A}}_{j0} - \mathbf{A}_{j0}\r)  = 0.
\end{eqnarray}
Consequently, the CMB signal affects neither the best-fit values of the spectral parameters, $\beta$, nor their uncertainties.
 
\section{Residuals power spectrum and its variance.}

\label{app:psRes}

We calculate the power spectrum of the residuals up to the second order in $\delta \beta$. From Eq.~(\ref{eqn:resTaylorHarFinal})  we get,
\begin{eqnarray}
{C}^{\rm res}_{l} & \simeq & \frac{1}{2l+1}\sum_{m} \langle \mathbf{\tilde r}^{\rm cmb, \, \dagger}_j \, \mathbf{\tilde r}^{\rm cmb}_j\rangle
=   \frac{1}{2l+1}\sum_{m}  \bigg[ \mathbf{\tilde y}_j^\dagger  \mathbf{\tilde y}_j
      +  {\rm tr} \Big[ \boldsymbol{\Sigma}\,\mathbf{\tilde Y}_j^{\l(1\r)\, \dagger}  \,  \mathbf{\tilde Y}^{\l(1\r)}_j \Big]   
      +  \mathbf{\tilde y}_j^\dagger {\rm tr} \Big[ \mathbf{\tilde Y}_j^{\l(2\r)} \boldsymbol{\Sigma} \Big] 
         +     {\rm tr} \Big[ \mathbf{\tilde Y}_j^{\l(2\r) \, \dagger\, } \boldsymbol{\Sigma} \Big] \mathbf{\tilde y}_j \bigg] \nonumber\\
 & = &  \frac{1}{2l+1}\bigg[ \sum_{m} \mathbf{\tilde y}_j^\dagger  \mathbf{\tilde y}_j
      + {\rm tr} \Big[ \boldsymbol{\Sigma} \, \sum_{m} \mathbf{\tilde Y}^{\l(1\r)\,\dagger\,}_j  \mathbf{\tilde Y}_j^{\l(1\r)} \Big]
      + \sum_{m} \Big( \mathbf{\tilde y}_j^{\dagger\,} \mathbf{\tilde z}_j + \mathbf{\tilde z}_j^{\dagger\,} \mathbf{\tilde y}_j\Big)\bigg],
      \label{eqn:psRes}
\end{eqnarray}
where $j = \ell^2 +\ell+m$. From this, Eq.~(\ref{eqn:psResFinal}) follows.

To calculate the expression for the variance we first compute, (here $j$ and $j'$ correspond to the same multipole, $\ell$, and two different values of $m$),
\begin{eqnarray}
\sum_{m,m'}  \langle  \mathbf{\tilde r}^{\rm cmb\, \dagger\,}_{j} \mathbf{\tilde r}^{\rm cmb}_{j}\, \mathbf{\tilde r}^{\rm cmb\, \dagger\,}_{j'} \mathbf{\tilde r}^{\rm cmb}_{j'} \rangle
\hskip -2pt  = \hskip -4pt  
\sum_{m,m'}&& \hskip -19pt \bigg[ 
{\rm tr}\l[ \mathbf{\tilde Y}^{\l(1\r) }_{j} \boldsymbol{\Sigma} \mathbf{\tilde Y}^{\l(1\r)  \dagger\,}_{j}\r] \, {\rm tr}\l[ \mathbf{\tilde Y}^{\l(1\r) }_{j'} \boldsymbol{\Sigma} \mathbf{\tilde Y}^{\l(1\r) \dagger\,}_{j'}\r] 
+ 2\, {\rm tr}\l[ \mathbf{\tilde Y}^{\l(1\r) \dagger\,}_{j} \boldsymbol{\Sigma} \mathbf{\tilde Y}^{\l(1\r)}_{j'}\r] \, {\rm tr}\l[ \mathbf{\tilde Y}^{\l(1\r) \dagger\,}_{j'} \boldsymbol{\Sigma} \mathbf{\tilde Y}^{\l(1\r)}_{j}\r] 
\nonumber \\
&+ &\mathbf{\tilde y}^{\dagger}_j  \mathbf{\tilde y}^{\phantom{\dagger}}_j \, \mathbf{\tilde y}^{\dagger}_{j'}  \mathbf{\tilde y}^{\phantom{\dagger}}_{j'} \hskip -2pt
+  \mathbf{\tilde y}^{\dagger}_j  \mathbf{\tilde y}_j {\rm tr} \Big[ \boldsymbol{\Sigma}\,\mathbf{\tilde Y}^{\l(1\r) \dagger}_{j'} \mathbf{\tilde Y}^{\l(1\r)}_{j'}\Big] \hskip -2pt
+  \mathbf{\tilde y}^{\dagger}_j  \mathbf{\tilde y}_{j'} {\rm tr}\Big[ \boldsymbol{\Sigma} \, \mathbf{\tilde Y}^{\l(1\r) \dagger\,}_{j'} \mathbf{\tilde Y}^{\l(1\r)}_{j} \Big]  \nonumber\\
 &+ & \mathbf{\tilde y}^{\dagger}_{j'}  \mathbf{\tilde y}_j {\rm tr}\Big[ \boldsymbol{\Sigma} \,  \mathbf{\tilde Y}^{\l(1\r) \dagger}_{j} \mathbf{\tilde Y}^{\l(1\r)}_{j'}\Big]
+ \mathbf{\tilde y}^\dagger_{j'}  \mathbf{\tilde y}_{j'}{\rm tr}\Big[ \boldsymbol{\Sigma} \,  \mathbf{\tilde Y}^{\l(1\r) \dagger}_{j} \mathbf{\tilde Y}^{\l(1\r)}_{j}\Big] \hskip -2pt
 +   \mathbf{\tilde y}^\dagger_{j}  \mathbf{\tilde y}_{j} \mathbf{\tilde y}^\dagger_{j'} {\rm tr} \Big[\mathbf{\tilde Y}^{\l(2\r)}_{j'} \boldsymbol{\Sigma}\Big] \hskip -2pt \nonumber\\
&+ & \mathbf{\tilde y}^\dagger_{j}  \mathbf{\tilde y}_{j} \mathbf{\tilde y}_{j'} {\rm tr} \Big[\mathbf{\tilde Y}^{\l(2\r)}_{j'} \boldsymbol{\Sigma}\Big] \hskip -2pt
+  \mathbf{\tilde y}^\dagger_{j'}  \mathbf{\tilde y}_{j'} \mathbf{\tilde y}^\dagger_{j} {\rm tr} \Big[\mathbf{\tilde Y}^{\l(2\r)}_{j} \boldsymbol{\Sigma}\Big]\hskip -2pt
 +   \mathbf{\tilde y}^\dagger_{j'}  \mathbf{\tilde y}_{j'} \mathbf{\tilde y}_{j} {\rm tr} \Big[\mathbf{\tilde Y}^{\l(2\r)}_{j} \boldsymbol{\Sigma}\Big] \bigg], 
\end{eqnarray}
where we have retained contributions the lowest order in $\boldsymbol \delta$ separately for the terms related to the bias and the bias-free ones.
From this we now have,
\begin{eqnarray}
{\displaystyle {\rm Var} (C_\ell^{\rm res})} & {\displaystyle =} & {\displaystyle \frac{1}{(2\ell+1)^2}\sum_{m,m'}  \langle  \mathbf{\tilde r}^{\rm cmb\, \dagger\,}_{j} \mathbf{\tilde r}^{\rm cmb}_{j}\, \mathbf{\tilde r}^{\rm cmb\, \dagger\,}_{j'} \mathbf{\tilde r}^{\rm cmb}_{j'} \rangle
- (C^{\rm res}_\ell)^2}\nonumber \\ 
& {\displaystyle = }&  {\displaystyle \frac{1}{(2\ell+1)^2} \,\Bigg\{ 2\,\bigg[\sum_m{\rm tr}\l( \mathbf{\tilde Y}^{\l(1\r)}_{j} \boldsymbol{\Sigma} \mathbf{\tilde Y}^{\l(1\r)  \dagger\,}_{j}\r)\bigg]^2
+ \sum_{m,m'} \bigg[ \mathbf{\tilde y}^{\dagger}_j  \mathbf{\tilde y}_{j'} {\rm tr}\Big[ \boldsymbol{\Sigma} \, \mathbf{\tilde Y}^{\l(1\r) \dagger\,}_{j'} \mathbf{\tilde Y}^{\l(1\r)}_{j} \Big] \hskip -2pt
 +  \mathbf{\tilde y}^{\dagger}_{j'}  \mathbf{\tilde y}_j {\rm tr}\Big[ \boldsymbol{\Sigma} \,  \mathbf{\tilde Y}^{\l(1\r) \dagger}_{j} \mathbf{\tilde Y}^{\l(1\r)}_{j'}\Big]\bigg]\Bigg\},} \ \ \ \ \ \ \ 
\end{eqnarray}
which is equivalent to Eq.~(\ref{eqn:varResFinal}).

\section{Cosmological parameter likelihood}

\label{app:parLikeDervs}

Here, we derive explicit equations for the calculation of first order derivatives of the cosmological parameter likelihood. We note that the same type of calculations are in principle involved in the computation of the matrix of the second derivatives of the likelihood. Importantly, also these derivatives can be expressed solely via cross-spectra of the pixel-domain objects defined in Sect.~\ref{s2sect:compSepRes}. However, the number of terms of the analytic expression grows rapidly and we have found that in practice computing this matrix numerically is more efficient and thus useful, while ensuring precision sufficient for the purpose. For this reason we do not include explicitly the corresponding derivations. 

\subsection{Preliminaries}
Let us first compute the inverse of the covariance matrix, $\mathbf{C}$. This can be done with help of the Sherman-Morrison-Woodbury formula~(e.g., \cite{GolubvanLoan}),
\begin{equation}
	\centering
		\mathbf{C}^{-1} = \mathbf{D}^{-1} -  \mathbf{D}^{-1} \mathbf{\tilde Y}^{\l(1\r)} \underbrace{(\boldsymbol{\Sigma}^{-1} + \mathbf{\tilde Y}^{\l(1\r) \dagger}
		\mathbf{D}^{-1} \mathbf{\tilde Y}^{\l(1\r)})^{-1}}_{\displaystyle \equiv \mathbf{U}} \mathbf{\tilde Y}^{\l(1\r) \dagger}\mathbf{D}^{-1},
		\label{eqn:cMatinv}
\end{equation}
where $\mathbf{U}$ is a square, real, symmetric matrix of rank $n_\beta$.
On noting that,
\begin{eqnarray}
[\mathbf{\tilde Y}^{\l(1\r) \dagger} \mathbf{D}^{-1} \mathbf{\tilde Y}^{\l(1\r)} ]_{\beta \beta'} = \sum_{j,j'} \mathbf{\tilde Y}^{\l(1\r) \dagger}_{j \beta}(\mathbf{D}^{-1})_{j j'}   \mathbf{\tilde Y}^{\l(1\r)}_{j' \beta'} 
= \sum_\ell C_\ell^{-1} (2\ell+1) \, {\botimes}_\ell( \mathbf{\tilde Y}^{\l(1\r)}_\beta, \mathbf{\tilde Y}^{\l(1\r)}_{\beta'}),
\end{eqnarray}
where $ j = \ell^2+\ell+m$, the inverse of $\mathbf{U}$ can be expressed as,
\begin{eqnarray}
\mathbf{U^{-1}} = \boldsymbol{\Sigma}^{-1} + \mathbf{\tilde Y}^{\l(1\r)\, \dagger}
		\mathbf{D}^{-1} \mathbf{\tilde Y}^{\l(1\r)}  = \boldsymbol{\Sigma}^{-1} + \sum_\ell C_\ell^{-1} (2\ell+1) \, {\botimes}_\ell( \mathbf{\tilde Y}^{\l(1\r)}, \mathbf{\tilde Y}^{\l(1\r)}).
\label{eqn:invU}
\end{eqnarray}

Typically, this matrix will be dense and its inversion has to be then calculated numerically. However, given the typically limited number of spectral parameters, 
this does not pose any computational problems and matrix $\mathbf{U}$ can be readily derived. 

We can now write down the explicit expression for $(\mathbf{C}^{-1})_{jj'}$, which is given by,
\begin{eqnarray}
(\mathbf{C}^{-1})_{jj'} & = & C_\ell^{-1} \delta_{jj'} - \sum_{j'', j''' \atop \beta, \beta'} C_\ell^{-1} \delta_{jj''} \mathbf{\tilde Y}^{\l(1\r)}_{j''\beta}\, \mathbf{U}_{\beta \beta'} \mathbf{\tilde Y}^{\l(1\r) \dagger}_{j'''\beta'}\, C_{\ell'}^{-1} \delta_{j''' j'} 
= C_\ell^{-1} \delta_{jj'} -  C_\ell^{-1} C_{\ell'}^{-1} \sum_{\beta, \beta'}  \mathbf{\tilde Y}^{\l(1\r)}_{j\beta}\, \mathbf{U}_{\beta \beta'} \mathbf{\tilde Y}^{\l(1\r) \dagger}_{j'\beta'}, \ \ \ \ 
\label{eqn:invCelem}
\end{eqnarray}
where $\ell$ is related to $j$ and $\ell'$ to $j'$. Given this we can write,
\begin{eqnarray}
(\mathbf{C}^{-1})_{jj'}^2 & = & C_\ell^{-2} \delta_{jj'} -  2C_\ell^{-3} \delta_{j j'} \sum_{\beta, \beta'}  \mathbf{\tilde Y}^{\l(1\r)}_{j\beta} \mathbf{U}_{\beta \beta'} \mathbf{\tilde Y}^{\l(1\r) \dagger}_{j'\beta'}
+  C_\ell^{-2} C_{\ell'}^{-2}
 \sum_{\beta, \beta' \atop \alpha, \alpha'}  \mathbf{\tilde Y}^{\l(1\r)}_{j\beta} \mathbf{\tilde Y}^{\l(1\r)}_{j\alpha} \mathbf{U}_{\beta \beta'} \mathbf{U}_{\alpha \alpha'} \mathbf{\tilde Y}^{\l(1\r) \dagger}_{j'\beta'} \mathbf{\tilde Y}^{\l(1\r)\dagger}_{j'\alpha'}, \ \ \ \  \ \ \ \ \ \ \  \  \ \ 
 \label{eqn:cij_to_-2}
\end{eqnarray}
which will be found useful later on.

Similarly, for an arbitrary harmonic space vector, $\mathbf{\tilde x}$, we have on using Eq.~(\ref{eqn:invCelem}),
\begin{eqnarray}
(\mathbf{C}^{-1} \mathbf{\tilde x})_{j}  & = & 
 \equiv   C_\ell^{-1} \Big( \mathbf{\tilde x}_{j} - \hskip -2pt \sum_\beta \, \mathbf{\tilde Y}^{\l(1\r)}_{j \beta}\, \mathbf{u}_\beta[\mathbf{\tilde x}]\Big),
 \label{eqn:invCxVect}
\end{eqnarray}
where,
\begin{eqnarray}
\mathbf{u}_\beta[\mathbf{\tilde x}] \equiv \sum_{\beta'}\,\mathbf{U}_{\beta\beta'} \,  \sum_{\ell'} \frac{2\ell'+1}{C_{\ell'}} \, \botimes_{\ell'}\hskip -2pt( \mathbf{\tilde Y}_{\beta'}^{\l(1\r)},  \mathbf{\tilde x}).
\end{eqnarray}

\subsection{Likelihood}

The '$\chi^2$-term' of the likelihood can be represented as follows, Eq.~\eref{eqn:parLikeAv},
\begin{eqnarray}
{\rm tr}\, \mathbf{C}^{-1}\mathbf{E} = {\rm tr}\, \mathbf{C}^{-1}\mathbf{\hat C} + {\rm tr}\, \mathbf{C}^{-1}\l(\mathbf{E}-\mathbf{\hat C}\r),
\label{eqn:parLikeChi2term}
\end{eqnarray}
where $\mathbf{\hat C}$ stands for the true signal covariance matrix. We can write the first term on the rhs of this equation as,
\begin{eqnarray}
{\rm tr}\, \mathbf{C}^{-1}\mathbf{\hat C} = \sum_{j, j'} (\mathbf{C}^{-1})_{jj'} \mathbf{\hat C}_{j'j} & = & \sum_{j,j'}
\Big(C_\ell^{-1} \delta_{jj'} -  C_\ell^{-1} C_{\ell'}^{-1} \sum_{\beta, \beta'}  \mathbf{\tilde Y}^{\l(1\r)}_{j\beta}\, \mathbf{U}_{\beta \beta'} \mathbf{\tilde Y}^{\l(1\r) \dagger}_{j'\beta'}\Big)
\Big(\hat{C}_{\ell'} \delta_{jj'} +   \sum_{\beta,\beta'}\, \mathbf{\tilde Y}^{\l(1\r)}_{j\beta} \boldsymbol{\Sigma}_{\beta\beta'} \mathbf{\tilde Y}^{\l(1\r) \dagger}_{j'\beta'}\Big)\nonumber\\
& = & 
\sum_{\ell}\,\bigg[(2\ell+1) \frac{ \hat{C}_{\ell}}{C_\ell}\, \Big(1 - C_{\ell}^{-1}\, {\rm tr}\, \big[\mathbf{U}\, \botimes_\ell\hskip -2pt( \mathbf{\tilde Y}^{\l(1\r)}, \mathbf{\tilde Y}^{\l(1\r)})\big]\Big)
+ \frac{(2\ell+1)}{C_{\ell}}\,{\rm tr}\, \big[\boldsymbol{\Sigma}\, \botimes_\ell\hskip -2pt( \mathbf{\tilde Y}^{\l(1\r)}, \mathbf{\tilde Y}^{\l(1\r)})\big] \bigg]\nonumber\\
& - & \sum_{\ell, \ell'} \, \frac{(2\ell+1)}{C_\ell} \, \frac{(2\ell'+1)}{C_{\ell'}} {\rm tr}\, \big[ \mathbf{U} \, \botimes_{\ell'}\hskip -2pt( \mathbf{\tilde Y}^{\l(1\r)}, \mathbf{\tilde Y}^{\l(1\r)})\, \boldsymbol{\Sigma}\, \botimes_{\ell}\hskip -2pt( \mathbf{\tilde Y}^{\l(1\r)}, \mathbf{\tilde Y}^{\l(1\r)})\big],
\label{eqn:parLikeTerm0}
\end{eqnarray}
where for the computational reasons it is better to first performs the sum over $\ell$ and $\ell'$ before calculating traces. In particular the last term becomes then linear in $\ell_{max}$.
Similarly, we can now write the second term of the $\chi^2$,
\begin{eqnarray}
{\rm tr}\, \mathbf{C}^{-1}\l(\mathbf{E}-\mathbf{\hat C}\r) & = & \sum_{j,j'} (\mathbf{C}^{-1})_{jj'} \big(\mathbf{\tilde y}_{j'} \mathbf{\tilde y}_{j}^\dagger + \mathbf{\tilde z}_{j'} \mathbf{\tilde y}_j^\dagger + \mathbf{\tilde y}_{j'} \mathbf{\tilde z}_j^\dagger \big)
 =  \sum_\ell  \frac{(2\ell+1)}{C_\ell} \big(\botimes_\ell\hskip -2pt(\mathbf{\tilde y}, \mathbf{\tilde y}) + \botimes_\ell\hskip -0pt(\mathbf{\tilde z}, \mathbf{\tilde y}) + \botimes_\ell\hskip -0pt(\mathbf{\tilde y}, \mathbf{\tilde z})\big)
\nonumber\\
& - & \sum_{\ell, \ell'}\,\frac{(2\ell+1)}{C_\ell} \, \frac{(2\ell'+1)}{C_{\ell'}} \, {\rm tr} \,  \Big[ \mathbf{U}\, \Big(\botimes_{\ell'}\hskip -2pt(\mathbf{\tilde Y}^{\l(1\r)}, \mathbf{\tilde y})\, \botimes_\ell\hskip -2pt(\mathbf{\tilde y}, \mathbf{\tilde Y}^{\l(1\r)}) + \botimes_{\ell'}\hskip -1pt(\mathbf{\tilde Y}^{\l(1\r)}, \mathbf{\tilde y})\, \botimes_\ell\hskip -2pt(\mathbf{\tilde z}, \mathbf{\tilde Y}^{\l(1\r)}) \nonumber\\
& & \hskip 5truecm + \botimes_{\ell'}\hskip -2pt(\mathbf{\tilde Y}^{\l(1\r)}, \mathbf{\tilde z})\, \botimes_\ell\hskip -2pt(\mathbf{\tilde y}, \mathbf{\tilde Y}^{\l(1\r)})\Big)\Big],
\label{eqn:parLikeTerm1}
\end{eqnarray}
and again in a numerical implementation it is better to first perform sums over the multipoles and only later  take the traces.

The determinant of the assumed covariance matrix, $\mathbf{C}$, can then be efficiently calculated by noting that,
\begin{eqnarray}
{\rm det} \, \l(\l[
\begin{array}{l l}
{\displaystyle \mathbf{D}} & {\displaystyle -\mathbf{\tilde Y}^{\l(1\r)}}\\
{\displaystyle \mathbf{\tilde Y}^{\l(1\r)\, \dagger}} & {\displaystyle \phantom{-}\boldsymbol{\Sigma}^{-1}}
\end{array}
\r]\r)
= {\rm det} \, \mathbf{D} \, {\rm det}\, \big( \boldsymbol{\Sigma}^{-1} + \mathbf{\tilde Y}^{\l(1\r)\, \dagger}\,  \mathbf{D}^{-1} \, \mathbf{\tilde Y}^{\l(1\r)} \big) 
= {\rm det}\,  \boldsymbol{\Sigma}^{-1} \, {\rm det}\, \big( \mathbf{D} + \mathbf{\tilde Y}^{\l(1\r)}\,  \boldsymbol{\Sigma} \, \mathbf{\tilde Y}^{\l(1\r)\, \dagger} \big),
\end{eqnarray}
and thus
\begin{eqnarray}
{\rm det} \, \mathbf{C} = {\rm det} \, \mathbf{D} \, \frac{ {\rm det}\, \boldsymbol{\Sigma}}{{\rm det}\, \mathbf{U}}.
\label{eqn:parLikeDetTerm}
\end{eqnarray}
Given that $\mathbf{D}$ is diagonal and $\mathbf{U}$ and $\boldsymbol{\Sigma}$ -- small, the computation of the determinant of the covariance does not pose typically any problem. Note that as these two latter matrices are typically dense their determinants need to be calculated numerically using standard techniques.

\subsection{First derivatives of the likelihood}

We can now compute the likelihood gradient. From Eq.~\eqref{eqn:parGrad0} we can write,
\begin{eqnarray}
 \langle {\cal S}^{par}_{,i}\rangle  & = & {\rm tr} \l[\mathbf{C}^{-1} \mathbf{C}_{,i}\, - \,\mathbf{C}^{-1} \mathbf{C}_{,i}\,\mathbf{C}^{-1} \,\mathbf{E}\r] 
 					= {\rm tr} \l[\mathbf{C}^{-1} \mathbf{C}_{,i}\, - \,\mathbf{C}^{-1} \mathbf{C}_{,i}\,\mathbf{C}^{-1} \,\mathbf{\hat C}\r] - {\rm tr} \l[\mathbf{C}^{-1} \mathbf{C}_{,i}\,\mathbf{C}^{-1} (\mathbf{E}-\mathbf{\hat C})\r],
\label{eqn:gradInit}
 \end{eqnarray}
 where the second part contains all the extra corrections from the model mismatches, while the former vanishes when the true and estimated parameters are the same, i.e, $\mathbf{\hat C} = \mathbf{C}$.

 We compute each of these two terms separately below.
 
 \subsubsection{${\rm tr} \l[\mathbf{C}^{-1} \mathbf{C}_{,i}\, - \,\mathbf{C}^{-1} \mathbf{C}_{,i}\,\mathbf{C}^{-1} \,\mathbf{\hat C}\r]:$}
 We first observe that,
 \begin{eqnarray}
\mathbf{\hat C} -  \mathbf{C} =  \mathbf{\hat D} -  \mathbf{D}, 
 \end{eqnarray}
 where $\mathbf{\hat D}$ stands for $\mathbf{D}$ computed for the true values of the cosmological parameters. Hence, we can represent $\mathbf{\hat C}$ as,
 \begin{eqnarray}
 \mathbf{\hat C} = \mathbf{C} + \mathbf{\hat D} -  \mathbf{D} \equiv \mathbf{C} + \Delta\mathbf{D},
 \label{eqn:trueCMBcorr}
 \end{eqnarray}
 where,
 \begin{eqnarray}
 \Delta\mathbf{D}_{jj'} = ({\hat C}_\ell-C_\ell)\, \delta_{jj'} \equiv \Delta C_\ell^{\rm cmb} \, \delta_{jj'}.
 \end{eqnarray}
 On using Eq.~(\ref{eqn:trueCMBcorr}) we can rewrite the complete term as,
 \begin{eqnarray}
 {\rm tr} \l[\mathbf{C}^{-1} \mathbf{C}_{,i}\, - \,\mathbf{C}^{-1} \mathbf{C}_{,i}\,\mathbf{C}^{-1} \,\mathbf{\hat C}\r] = -{\rm tr} \l[\mathbf{C}^{-1} \mathbf{C}_{,i}\,\mathbf{C}^{-1} \,\Delta\mathbf{D}\r],
 \end{eqnarray}
and then,
 \begin{eqnarray}
&& \hskip -20pt{\rm tr} \l[\mathbf{C}^{-1} \mathbf{C}_{,i}\,\mathbf{C}^{-1} \,\Delta\mathbf{D}\r]  =  \sum_{j,j',j'',j'''} \, (\mathbf{C}^{-1})_{jj'}\, (\mathbf{C}_{,i})_{j'j''}\, (\mathbf{C}^{-1})_{j''j'''} \, \Delta\mathbf{D}_{j'''j}
 = \sum_{j,j'}   (\mathbf{C}^{-1})_{jj'}^2 \frac{\partial C^{\rm cmb}_{\ell'}}{\partial p_i} \, \Delta C^{\rm cmb}_\ell
\nonumber \\
& & \hskip 0pt = \sum_{j,j'} \Big(C_\ell^{-2} \delta_{jj'} -  2C_\ell^{-3} \delta_{j j'} \sum_{\beta, \beta'}  \mathbf{\tilde Y}^{\l(1\r)}_{j\beta}\, \mathbf{U}_{\beta \beta'} \mathbf{\tilde Y}^{\l(1\r) \dagger}_{j'\beta'}
+  \, C_\ell^{-2} C_{\ell'}^{-2}
 \sum_{\beta, \beta' \atop \alpha, \alpha'}  \mathbf{\tilde Y}^{\l(1\r)}_{j\beta}\, \mathbf{\tilde Y}^{\l(1\r)}_{j\alpha} \, \mathbf{U}_{\beta \beta'} \, \mathbf{U}_{\alpha \alpha'} \, \mathbf{\tilde Y}^{\l(1\r) \dagger}_{j'\beta'}\, \mathbf{\tilde Y}^{\l(1\r)  \dagger}_{j'\alpha'}\Big)
  \frac{\partial C_{\ell'}^{\rm cmb}}{\partial p_i}  \Delta C_\ell,\nonumber\\
& & =  \sum_{\ell}
\frac{\partial C^{\rm cmb}_{\ell}}{\partial p_i} \, \Delta C_{\ell} \frac{(2\ell+1)}{C_\ell^2}
\, -  \, {\rm tr} \Big\{ \mathbf{U}\, \sum_\ell \Big[ \frac{2(2\ell+1)}{C_\ell^3}\, \frac{\partial C^{\rm cmb}_{\ell}}{\partial p_i} \, \Delta C_{\ell} \, \botimes_\ell(\mathbf{\tilde Y}^{\l(1\r)}, \mathbf{\tilde Y}^{\l(1\r)})\Big] \Big\}
+ \nonumber\\
& &  \hskip 50pt+ 
{\rm tr}\Big\{ \mathbf{U} \, \sum_\ell \Big[  \frac{(2\ell+1)}{C_\ell^{2}} \frac{\partial C^{\rm cmb}_{\ell}}{\partial p_i} \botimes_\ell(\mathbf{\tilde Y}^{\l(1\r) \dagger}, \mathbf{\tilde Y}^{\l(1\r)})\Big] \mathbf{U}  
\sum_\ell \Big[ \frac{(2\ell+1)}{C_{\ell}^{2}} \, \Delta C_{\ell} \, \botimes_{\ell}\hskip -2pt(\mathbf{\tilde Y}^{\l(1\r)}, \mathbf{\tilde Y}^{\l(1\r) \dagger})\Big] \Big\}
\label{eqn:gradTerm00}
 \end{eqnarray}
where $p_i$ is the parameter we estimate and we have used Eq.~(\ref{eqn:cij_to_-2}) derived earlier and the fact that $\mathbf{U}$ is positive and symmetric. We note that the expression on the right hand side above is manifestly real as it should.
 
 \subsubsection{${\rm tr} \l[\mathbf{C}^{-1} \mathbf{C}_{,i}\,\mathbf{C}^{-1} (\mathbf{E}-\mathbf{\hat C})\r]:$}
 On using  Eq.~(\ref{eqn:EdefMat}) and the fact that the covariance, $\mathbf{C}$, is symmetric and real, we can rewrite this term as follows,
 \begin{eqnarray}
 \nonumber \\
 {\rm tr} \l[\mathbf{C}^{-1} \mathbf{C}_{,i}\,\mathbf{C}^{-1} (\mathbf{E}-\mathbf{\hat C})\r]
 & = &  
 {\rm tr} \bigg\{ \mathbf{C}_{,i}\, \Big[(\mathbf{C}^{-1} \mathbf{\tilde y}) (\mathbf{C}^{-1} \mathbf{\tilde y})^\dagger + (\mathbf{C}^{-1} \mathbf{\tilde y}) (\mathbf{C}^{-1} \mathbf{\tilde z})^\dagger + (\mathbf{C}^{-1} \mathbf{\tilde z}) (\mathbf{C}^{-1} \mathbf{\tilde y})^\dagger\Big] \bigg\}
\nonumber \\
& = & \sum_{j} \bigg\{ \frac{\partial C_\ell}{\partial p_i} \Big[ (\mathbf{C}^{-1} \mathbf{\tilde y})_j (\mathbf{C}^{-1} \mathbf{\tilde y})_{j}^\dagger
+  (\mathbf{C}^{-1} \mathbf{\tilde z})_j (\mathbf{C}^{-1} \mathbf{\tilde y})_{j}^\dagger + (\mathbf{C}^{-1} \mathbf{\tilde y})_j (\mathbf{C}^{-1} \mathbf{\tilde z})_{j}^\dagger
\Big]\bigg\}.
\label{eqn:parGrad}
\end{eqnarray}
where as usual $j = \ell^2 + \ell + m$. Therefore, on using Eq.~(\ref{eqn:invCxVect}),
\begin{eqnarray}
{\rm tr} \l[\mathbf{C}^{-1} \mathbf{C}_{,i}\,\mathbf{C}^{-1} (\mathbf{E}-\mathbf{\hat C})\r]
\hskip -2pt 
& = & \hskip -2pt \sum_{\ell} \bigg\{ \frac{2\ell+1}{C_\ell^2} \, \frac{\partial C_\ell}{\partial p_i} \Big[ \botimes_\ell(\mathbf{\tilde y}, \mathbf{\tilde y}) - 2 \, \Re \sum_\beta \botimes_\ell( \mathbf{\tilde y}, \mathbf{\tilde Y}^{\l(1\r)}_{\beta})\, \mathbf{\tilde u}_\beta[\mathbf{\tilde y}]
+ \sum_{\beta, \beta'}\, \botimes_\ell( \mathbf{\tilde Y}^{\l(1\r)}_{\beta},  \mathbf{\tilde Y}^{\l(1\r)}_{\beta'}) \, \mathbf{\tilde u}^\dagger_\beta[\mathbf{\tilde y}] \, \mathbf{\tilde u}_{\beta'}[\mathbf{\tilde y}] 
\nonumber\\
&& \hskip -80pt 
+ 2 \, \Re \Big(\hskip -2pt\botimes_\ell(\mathbf{\tilde y}, \mathbf{\tilde z})  \hskip -2pt
- \hskip -2pt  \sum_\beta \botimes_\ell( \mathbf{\tilde y}, \mathbf{\tilde Y}^{\l(1\r)}_{\beta})\, \mathbf{\tilde u}_\beta[\mathbf{\tilde z}] - \sum_\beta \botimes_\ell(  \mathbf{\tilde z}, \mathbf{\tilde Y}^{\l(1\r)}_{\beta})
\, \mathbf{\tilde u}_\beta[\mathbf{\tilde y}]
 + \sum_{\beta, \beta'}\, \botimes_\ell( \mathbf{\tilde Y}^{\l(1\r)}_{\beta},  \mathbf{\tilde Y}^{\l(1\r)}_{\beta'}) \, \mathbf{\tilde u}^\dagger_\beta[\mathbf{\tilde y}] \, \mathbf{\tilde u}_{\beta'}[\mathbf{\tilde z}]\Big) 
\Big].
\label{eqn:gradTerm2}
\end{eqnarray}

\section{Special case: homogeneous noise and pixel-independent scaling laws.}

\label{app:simpCase}

Let us assume that the noise of the frequency maps is homogeneous and that we use global scaling laws for all considered components and all considered pixels. Then, matrix $\mathbf{W}_p^{0k}(\beta)$ and its derivatives with respect to the spectra parameters are all the same for all considered pixels, $p$.
We can then drop the pixel subscripts and introduce following pixel-independent objects,
\begin{eqnarray}
\mathbf{w}_k \equiv \mathbf{W}^{0k}(\bar \beta), \ \ \  \partial \mathbf{W}_{k \beta} \equiv \frac{\partial \mathbf{W}^{0k}}{\partial \beta}\bigg|_{\bar \beta},
\ \ \ \ \boldsymbol{v}_k \equiv \sum_{\beta,\beta'} \left.\frac{\partial^2 \mathbf{W}^{0k}}{\partial \beta\,\partial \beta'}\right|_{\bar \beta} {\boldsymbol \Sigma}_{\beta' \beta},
\end{eqnarray}
so then, we can rewrite Eqs.~\eref{eqn:yVectsDef} and~\eref{eqn:zFullPix} as,
\begin{eqnarray}
\mathbf{\tilde y} & = & \mathbf{\tilde F}\,\mathbf{w}\nonumber\\
\mathbf{\tilde Y^{\l(1\r)}} & = & \mathbf{\tilde F}\, \partial \mathbf{W}\\
\mathbf{\tilde z} & = & \mathbf{\tilde F}\,\boldsymbol{v}.\nonumber
\end{eqnarray}
Subsequently, we can relate all the cross-correlations appearing in the earlier equations to the
cross-correlation matrix of the foreground signal in different frequency bands, ${\cal F}^{\rm fore}_\ell$, defined in Eq.~(\ref{eqn:bandXspecDef}),
\begin{eqnarray}
\botimes_\ell(\mathbf{\tilde y}, \mathbf{\tilde y}) & = & \mathbf{w}^t {\cal F}^{\rm fore}_\ell \mathbf{w},\nonumber\\
\botimes_\ell(\mathbf{\tilde Y}^{\l(1\r)}, \mathbf{\tilde Y}^{\l(1\r)}) & = & \partial\mathbf{W}^t \, {\cal F}_\ell^{\rm fore}\, \partial\mathbf{W},\nonumber\\
\botimes_\ell(\mathbf{\tilde y}, \mathbf{\tilde z}) & = &  \mathbf{w}^t {\cal F}_\ell^{\rm fore} \boldsymbol{v},\\
\botimes_\ell(\mathbf{\tilde Y}^{\l(1\r)}, \mathbf{\tilde y}) & = & \partial\mathbf{W}^t \, {\cal F}_\ell^{\rm fore}\,\mathbf{w},\nonumber\\
\botimes_\ell(\mathbf{\tilde Y}^{\l(1\r)}, \mathbf{\tilde z}) & = & \partial\mathbf{W}^t \, {\cal F}_\ell^{\rm fore}\,\boldsymbol{v}.\nonumber
\end{eqnarray}

\end{widetext}

\bibliographystyle{apsrev}
\bibliography{newForecast}

\end{document}